\documentclass[11pt]{article}
\usepackage[margin=1in]{geometry}
\usepackage[T1]{fontenc}
\usepackage{lmodern}
\usepackage{amsmath,amssymb,amsthm,mathtools}
\usepackage{microtype,booktabs,array}
\usepackage[numbers,sort&compress]{natbib}
\usepackage[colorlinks=true,allcolors=blue]{hyperref}
\newcommand{\doi}[1]{doi:\,\href{https://doi.org/#1}{\nolinkurl{#1}}}

\newtheorem{theorem}{Theorem}[section]
\newtheorem{lemma}[theorem]{Lemma}
\newtheorem{corollary}[theorem]{Corollary}
\newtheorem{proposition}[theorem]{Proposition}
\theoremstyle{remark}
\newtheorem{remark}[theorem]{Remark}
\theoremstyle{definition}
\newtheorem{question}[theorem]{Question}

\newcommand{\Tr}{\operatorname{Tr}}
\newcommand{\tr}{\operatorname{tr}}
\newcommand{\ad}{\operatorname{ad}}
\newcommand{\E}{\mathbb E}
\newcommand{\Prob}{\mathbb P}
\newcommand{\R}{\mathbb R}
\newcommand{\C}{\mathbb C}
\newcommand{\Tcal}{\mathcal T}
\newcommand{\Gc}{\mathcal G_c}
\newcommand{\op}{\mathrm{op}}
\newcommand{\eps}{\epsilon}
\newcommand{\Lam}{\Lambda}
\newcommand{\Fock}{\mathcal F}

\title{Upper and lower bounds for few-copy Gaussian tomography,\\ and the absence of a fermionic warm regime}
\author{Ron Rubin\\Rubin Anders Scientific, Inc.}
\date{November 2026}

\begin{document}
\maketitle

\begin{abstract}
Chen, Mele, Mele and Preskill proved a temperature crossover for bosonic Gaussian tomography: cold states need $\Theta(m^3/\eps^2)$ copies under single-copy or few-copy measurements, warm ones only $\Theta(m^2/\eps^2)$. The author proved $\Omega(m^3/\eps^2)$ for adaptive single-copy tomography of mixed fermionic Gaussian states. This note shows that the crossover has no fermionic counterpart and extends the fermionic bound to measurements on blocks of $k$ copies. There is no warm regime: on the quadratic Gibbs balls $\beta\|h\|_F\le\beta_0$ of unknown Hamiltonian $h$, the single-copy rate is $\Theta(m^3/\eps^2)$ for every fixed $\beta_0>0$, at accuracies that resolve the ball, with a lower-bound constant independent of $\beta_0$. At the maximally mixed state a replicated-parameter Fisher budget gives $N>m^3/(8{,}553{,}600\,k\,\eps^2)$ for blocks of at most $k$ copies and all $\eps\le1/3600$; an exact formula for the maximal Fisher information of a measurement on $t$ copies, through the spinor tensor powers of $\mathfrak{so}(2m)$, and a Casimir argument show that the budget $mt^2$ behind it is attained asymptotically for every fixed block size, so the dependence $1/k$, against $1/\sqrt k$ near the vacuum, is a property of that budget; whether $1/k$ is the true rate is open. Near the vacuum, block protocols need $\Omega(m^3/(\eps^2\min\{\sqrt k,m\}))$ copies, with no logarithm, by the batching argument of Chen, Mele, Mele and Preskill on the exterior-power embedding of qudit tomography; parity superselection changes none of these rates, and particle-number superselection costs nothing on the number-conserving families. The embedding also carries upper bounds: a particle-number measurement followed by the $k$-copy qudit protocol of Nayak and Zhou attains the lower bound for every $k$ on the number-conserving fermionic Gaussian states with bounded fugacity trace and on the passive bosonic Gaussian states with bounded photon number, and a heterodyne stage that estimates and undoes displacement and squeezing extends this to all cold bosonic Gaussian states. This closes the $k$-entangled gap of Chen, Mele, Mele and Preskill on their cold class; intermediate temperatures remain open.
\end{abstract}

\section{Introduction}\label{sec:intro}

A Gaussian state on $m$ modes is determined by its first and second moments, and learning it to trace distance $\eps$ takes $\Theta(m^2/\eps^2)$ copies, for bosons \cite{CFGLMWW26,CMMP26} and for fermions \cite{WW25,CFGLMWW26}; the optimal protocols measure all copies jointly. Section~6 of \cite{CFGLMWW26} asked whether such entangled measurements are fundamentally required. For pure states they are not \cite{CFGLMWW26,BMEL25,CMFLMHCP26}. For mixed states the answer is yes, and it was established in two preprints posted three days apart in September 2026.

Chen, Mele, Mele and Preskill \cite{CMMP26} settled the bosonic case completely. Under the promise $\Sigma\succeq(\tfrac12+\nu)I$ on the covariance matrix, with $\nu$ the number of thermal photons per mode, the single-copy sample complexity is $\Theta(m^2\min\{m,1+\nu^{-1}\}/\eps^2)$, attained by heterodyne detection; cold states ($\nu\lesssim1/m$) are $m^3$-hard even for measurements on blocks of $k$ copies, with the lower bound $\Omega(m^3/(\eps^2\min\{\sqrt k,m\}))$, while warm states ($\nu\gtrsim1$) are learned at the classical rate $m^2$. They call the latter \emph{classical to learn}, and they ask (their Section~1.4) how this crossover generalises to finite-dimensional systems.

The author \cite{Rub26} proved the fermionic case for single-copy measurements: any adaptive protocol that measures one copy at a time, with arbitrary POVMs, fresh ancillas and classical feedforward but no quantum memory between copies, needs $N>m^3/(8{,}553{,}600\,\eps^2)$ copies to learn mixed fermionic Gaussian states to trace distance $\eps\le1/3600$, and on the class $\Gc$ of covariance matrices with $\|\Gamma\|_\op\le1-c$ this is tight, $\Theta(m^3/\eps^2)$, by matchgate shadows. The hard family sits at the maximally mixed state $I/2^m$, and the mechanism is a dimension-free Fisher-information budget: every one-copy measurement has Fisher trace at most $2m$ about the $m(2m-1)$ Gaussian parameters, uniformly on a ball around $I/2^m$. The same paper gave, in its Section~9, a second mechanism for the bosonic case near the vacuum: an embedding of qudit tomography into passive Gaussian states by second quantisation, which transports the qudit separation $d^3$ versus $d^2$ \cite{HHJWY17,OW16,CHLLS22,LN25} through the block-model bound of Keskin, Luo, Majid and Radzihovsky \cite{KLMR26}, at the price of a factor $\sqrt{\log(m/\eps)}$; Remark~9.6 there identified the weighted form of the block bound, Eq.~(40) of \cite{KLMR26}, as the way to remove the logarithm, and Remark~9.7 noted that exterior powers in place of symmetric powers give the same construction for fermions.

The same embedding was found independently in \cite{CMMP26}, Section~4, where it is called the fugacity-matrix encoding, and there the logarithm is removed: their Lemma~1 states the weighted block bound in exactly the form Remark~9.6 of \cite{Rub26} asked for, and their batching argument (Theorem~4 there) handles the random block sizes that the embedding produces. The present note takes their Lemma~1 as its one black box and does five things; the first two concern the maximally mixed state, where the embedding is unavailable, the next two the embedding itself, and the last the superselection rules.

\begin{enumerate}
\item[(i)] \emph{No warm regime.} The hard family of the fermionic bound of \cite{Rub26}, and of its $k$-copy extension in (ii), consists of quadratic Gibbs states $\propto\exp(-\beta H)$ with $\beta\|h\|_F\le120\sqrt3\,\eps$, which is the high-temperature limit of every such family. Theorem~\ref{prop:nowarm} makes this precise: for every $\beta_0>0$ the class of quadratic Gibbs states with $\beta\|h\|_F\le\beta_0$ has single-copy rate $\Theta(m^3/\eps^2)$ on the accuracy window $\eps\le\min\{1/3600,\beta_0/(120\sqrt3)\}$, with a lower-bound constant independent of $\beta_0$. Raising the temperature shrinks the window linearly but never lowers the rate. This answers the question of \cite[Sec.~1.4]{CMMP26} for fermionic Gaussian states: there is no temperature at which they become classical to learn, because the fermionic covariance is bounded, $\|\Gamma\|_\op\le1$, and nothing plays the role of the thermal-photon number $\nu\to\infty$.
\item[(ii)] \emph{At the maximally mixed state, fermions, $k$ copies, explicit constants.} The near-vacuum sampler of Section~\ref{sec:vacuum} gives no bounded-cost reduction from qudit tomography there, so the argument is a direct information bound. Theorem~\ref{thm:kcopy-mixed} extends the Fisher-budget argument of \cite{Rub26} to blocks: $t$ copies of a Gaussian state on $m$ modes are one Gaussian state on $mt$ modes with the parameter replicated, so the $mt$-mode budget bounds every joint measurement on the block by $2mt^2$. The result is $N>m^3/(8{,}553{,}600\,k\,\eps^2)$ for all $\eps\le1/3600$ and every $k\ge1$, with the dependence $1/k$ rather than the $1/\sqrt k$ of (iii). The $1/k$ is not an artefact of a loose budget: Theorem~\ref{thm:exact} computes the supremum of the Fisher information of a measurement on $t$ copies at the maximally mixed state exactly, as the dimension-weighted average of the squared highest weight over the decomposition of the $t$-th tensor power of the spinor representation of $\mathfrak{so}(2m)$; for $t=2$ this equals $4m(1-\binom{2m}{m}4^{-m})$, and Theorem~\ref{thm:asymptotic} shows by a Casimir-operator argument that $F_{\max}(m,t)\ge mt^2(1-\sqrt{2(t-1)/m}-3/m)$ for every $t$, so the budget $mt^2$ is attained asymptotically for every fixed block size. So blocks of $t\ll m$ copies gain a factor $(1-o(1))\,t$ per copy at infinite temperature, against $\sqrt t$ near the vacuum and $\sqrt t$ for unstructured states \cite{KLMR26}; the uniform Fisher-trace budget on which the proof rests is therefore sharp up to constants for $t\le m/32$, and a lower bound with the dependence $1/\sqrt k$ would need information beyond it. Whether the rate $m^3/k$ is attained is left open. The same argument with the quantum Fisher information in place of the measurement Fisher information gives the collective lower bound $N>m^2/(4{,}276{,}800\,\eps^2)$ at the maximally mixed state, for every protocol whatsoever (Theorem~\ref{thm:collective}); the two per-copy budgets, $2m$ for a single copy and $p=m(2m-1)$ for the quantum Fisher information, differ by exactly the factor $m$ that separates the two rates.
\item[(iii)] \emph{Near the vacuum, fermions, $k$ copies, no logarithm.} Theorem~\ref{thm:vacuum} runs the batching argument of \cite{CMMP26} on the exterior-power sampler of \cite[Rem.~9.7]{Rub26}. Every constant that changes is recomputed; the trial success probability $\det(I+w/2)/2\ge3/4$ happens to equal the bosonic one, and the readout constant improves from $16$ to $6.6$.
\item[(iv)] \emph{The $k$-copy rate of the cold classes, for every $k$.} Theorem~\ref{cor:kcopy-rate} shows that the lower bound of (iii), and the bosonic one of \cite{CMMP26}, are attained for every $k\ge1$ by a particle-number measurement followed, on the one-particle sector, by the $k$-copy qudit protocol of Nayak and Zhou \cite{NZ26}, which is proved for every block size and every accuracy, with the full-rank rate also obtained in \cite{PSW26,PSTW26}; and that the readout needs no fixed weight in the one-particle sector, since a state with little weight there is close to the vacuum and may be learned proportionally less accurately. The quantitative input is Lemma~\ref{lem:coldclass}(ii): $\|\rho_X-\rho_{X'}\|_1\le2\|X-X'\|_1$ for number-conserving fermionic Gaussian states with fugacity matrices $X,X'$, and $\le(2/\delta)\|X-X'\|_1$ for passive bosonic ones with fugacity eigenvalues at most $1-\delta$, uniformly in the number of modes and in the fugacity trace; the fermionic case is also a corollary of the trace-norm covariance bound of Bittel, Mele, Eisert and Leone \cite{BMEL25}. So the $k$-copy rate of the number-conserving fermionic Gaussian states with fugacity trace at most $C$, and of the passive bosonic Gaussian states with at most $C$ photons and fugacity eigenvalues at most $1-\delta$, is $\Theta(m^3/(\eps^2\min\{\sqrt k,m\}))$ for all $k$, with constants depending on $C$ and $\delta$. This closes the first open problem in the outlook of \cite{CMMP26}, ``closing the $k$-entangled gap'', on the passive part of their cold class $\Sigma\preceq(\tfrac12+O(1/m))I$. Theorem~\ref{thm:cold-general} then closes it on the whole cold class, with any displacement and squeezing: heterodyne detection on $O(m^2/\eps^2)$ copies estimates the mean and the squeezing factor of the covariance, both are undone on every further copy, and the readout runs on the result. Two lemmas make this work: the residual displacement and squeezing after the correction enter the trace distance through Frobenius norms (Lemma~\ref{lem:perturb}, a Mandelstam--Tamm-type bound $T(\rho,e^{-iH}\rho e^{iH})\le\operatorname{Var}_\rho(H)^{1/2}$ with the variance computed by Wick's theorem), and the pure-squeezer factor $P$ of a covariance $\Sigma=P\Sigma_pP$, $\Sigma_p$ passive, is the explicit matrix function $P^2=(-(\Sigma\Omega)^2)^{-1/2}\Sigma$, Lipschitz in Frobenius norm near the vacuum although the Williamson form is not (Lemma~\ref{lem:squeezer}). Intermediate temperatures remain open (Remark~\ref{rem:cold-general}).
\item[(v)] \emph{Superselection rules.} Physical fermionic measurements are even, and often number-conserving; Möbus, Stilck Fran\c ca and Rouz\'e \cite{MSR26}, who learn local fermionic Lindbladians under the parity superselection rule, ask which coefficients remain learnable with number-conserving access. For Gaussian-state tomography the answer is complete (Proposition~\ref{prop:superselection}): parity superselection changes no rate in the fermionic rows of Table~\ref{tab:rates}, and on the number-conserving Gaussian states, which form an $m^2$-dimensional slice of the family, both lower bounds at the maximally mixed state hold with the same constants and number superselection costs nothing, so the separation $\Theta(m^2)$ versus $\Theta(m^3)$ between collective and single-copy measurements persists there for covariance a fixed distance from the pure boundary.
\end{enumerate}

Table~\ref{tab:rates} collects the rates. Proposition~\ref{prop:transfer} states the embedding as a transfer lemma, so that other qudit lower bounds in the block model can be moved to Gaussian states near the vacuum without repeating the batching argument. Section~\ref{sec:open} lists what remains open, including the $k$-copy upper bounds on the general classes and the $\sqrt k$ at the maximally mixed state.

\begin{table}[t]
\centering\footnotesize
\begin{tabular}{@{}l@{\hskip 6pt}l@{\hskip 6pt}l@{\hskip 6pt}l@{}}
\toprule
Class of $m$-mode mixed Gaussian states & Single-copy & Blocks of $\le k$ copies & Collective\\
\midrule
Bosons, cold, $\Sigma\preceq(\tfrac12+c'/m)I$ \cite{CMMP26}, here & $\Theta(m^3)$ & $\Theta(m^3/\min\{\sqrt k,m\})^{\dagger}$ & $\Theta(m^2)$\\
Bosons, $\Sigma\succeq(\tfrac12+\nu)I$ \cite{CMMP26} & $\Theta(m^2\min\{m,1+\nu^{-1}\})$ & & $\Theta(m^2)$\\
Fermions, $\|\Gamma\|_\op\le1-c$, $0<c<1$ fixed \cite{Rub26}, here & $\Theta(m^3)$ & $\Omega(\max\{m^3/k,m^2\})$ & $\Theta(m^2)$\\
Fermions, number-conserving, $\tr X\le C$, here & $\Theta(m^3)$ & $\Theta(m^3/\min\{\sqrt k,m\})^{\dagger}$ & $\Theta(m^2)$\\
Fermions, all mixed \cite{Rub26}, here & $\Omega(m^3),\ O(m^4)$ & $\Omega(m^3/\min\{\sqrt k,m\})$ & $\Theta(m^2)$\\
\bottomrule
\end{tabular}
\caption{Copies, in units of $1/\eps^2$, to learn to trace distance $\eps$ with probability $2/3$, for $\eps$ below a constant and $m$ above a constant, which on the rows with a class parameter ($c$, $c'$ or $C$) may depend on that parameter, as may the implied constants; protocols may be adaptive, and block protocols keep no quantum memory between blocks. $\nu$ is the thermal-photon number per mode; $c'\ge\tfrac23$ the fixed bosonic coldness parameter; $c$ the gap of the fermionic covariance from the pure boundary; $C\ge\tfrac12$ a fixed bound on the fugacity trace. $\dagger$: Theorems~\ref{cor:kcopy-rate} and~\ref{thm:cold-general}, for every $k$: on the number-conserving fermionic states with fugacity trace at most $C$, and on all cold bosonic Gaussian states $\Sigma\preceq(\tfrac12+c'/m)I$ with any displacement and squeezing, where the lower bound needs $c'\ge\tfrac23$. The collective upper bounds are from \cite{CFGLMWW26,CMMP26}; the collective lower bounds on the fermionic classes are Theorem~\ref{thm:collective} (fixed $c$, all mixed) and the embedding with quantum memory \cite[Cor.~9.5]{Rub26} (near vacuum).}
\label{tab:rates}
\end{table}

\paragraph{Credit.} The bosonic results quoted here are entirely due to \cite{CMMP26}, and the removal of the logarithm in the near-vacuum reduction is their Lemma~1 and Theorem~4; this note contributes the fermionic transcription of that argument, with its constants, the transfer lemma, the $k$-copy rate of the cold number-conserving classes for every $k$ (Theorem~\ref{cor:kcopy-rate}, whose qudit ingredient is \cite{NZ26,PSW26,PSTW26}) and of the whole cold bosonic class (Theorem~\ref{thm:cold-general}), the results at the maximally mixed state (Theorems~\ref{thm:kcopy-mixed} and~\ref{thm:collective}), the absence of a warm regime (Theorem~\ref{prop:nowarm}) and the superselection statement (Proposition~\ref{prop:superselection}). The reduction itself was found in \cite{Rub26} and \cite{CMMP26} independently.

\section{Setting}\label{sec:setting}

We use the notation of \cite{Rub26}, recalled briefly. On $m$ modes, $c_1,\dots,c_{2m}$ are Majorana operators, $\{c_a,c_b\}=2\delta_{ab}I$, on the Fock space of dimension $d=2^m$; $B_{ab}=ic_ac_b$ for $a<b$ are the $p=m(2m-1)$ quadratic operators, orthonormal for the normalised trace $\tau=d^{-1}\Tr$. A state is fermionic Gaussian if it obeys Wick's theorem; it is determined by its covariance matrix $\Gamma_{ab}=\Tr(\rho\,ic_ac_b)$, a real antisymmetric $2m\times2m$ matrix with $\|\Gamma\|_\op\le1$, and $\Gc=\{\|\Gamma\|_\op\le1-c\}$. The full-rank family
\begin{equation}\label{eq:family}
K_\theta=\sum_{a<b}\theta_{ab}B_{ab},\qquad z_\theta=\tau(e^{K_\theta}),\qquad \rho_\theta=\frac{e^{K_\theta}}{d\,z_\theta},\qquad\theta\in\R^p,
\end{equation}
consists of even Gaussian states with $\rho_0=I/d$; $\rho_\theta$ is the Gibbs state $e^{-\beta H}/\Tr e^{-\beta H}$ of the quadratic Hamiltonian $H=\sum_{a<b}h_{ab}B_{ab}$ at inverse temperature $\beta$ when $\theta=-\beta h$, and we write $\|h\|_F=(\sum_{a<b}h_{ab}^2)^{1/2}$, so that $\|\theta\|_2=\beta\|h\|_F$; this is $1/\sqrt2$ times the Frobenius norm of the full antisymmetric coefficient matrix, which is the convention of \cite[Sec.~7]{Rub26}. By \cite[Lemma~2.1]{Rub26} there is a real orthogonal rotation of the Majoranas in which $K_\theta=\sum_{j=1}^m\lambda_jZ_j$ with commuting involutions $Z_j$ and $\lambda_j\ge0$, $\sum_j\lambda_j^2=\|\theta\|_2^2$; the normal-form values of $\Gamma(\rho_\theta)$ are $\tanh\lambda_j$ \cite[Eq.~(2.3)]{Rub26}. Throughout, $T(\rho,\sigma)=\tfrac12\|\rho-\sigma\|_1$.

\paragraph{The block model.} A $k$-copy protocol receives $N$ copies of an unknown state and, in round $j$, as a function of its classical history and internal randomness, chooses an integer $1\le t_j\le k$ and applies an arbitrary POVM, with fresh ancillas, to $t_j$ fresh copies; it keeps unlimited classical information and no quantum memory between rounds, may stop at a random time, and satisfies $\sum_jt_j\le N$ almost surely. This is the model of \cite[Def.~1.1]{KLMR26} and \cite[Sec.~4]{CMMP26}; $k=1$ is the single-copy model of \cite[Thm.~1.1]{Rub26}. In Section~\ref{sec:vacuum} we also use the following weighted form of the block bound, which is Lemma~1 of \cite{CMMP26}, a restatement of Eq.~(40), Proposition~7.1 and Corollary~6.2 of \cite{KLMR26}.

\begin{lemma}[{\cite[Lemma~1]{CMMP26}}]\label{lem:fd}
There are universal constants $\eps_{\rm fd}>0$ and $d_0\in\mathbb N$ such that the following holds for $d\ge d_0$ and $0<\eps<\eps_{\rm fd}$. Consider a protocol with a deterministic total copy budget $M$ that, in round $j$, applies a POVM to $t_j\ge1$ fresh copies of an unknown $\sigma\in\mathcal D(\C^d)$, where $t_j$ and the POVM are chosen before those copies are accessed, using only previous classical outcomes and internal randomness; no quantum memory is kept between rounds; the number of rounds $T$ may be random but $\sum_{j\le T}t_j\le M$ almost surely. If for every $\sigma\preceq2I/d$ the protocol outputs $\hat\sigma$ with $\Prob_\sigma(T(\hat\sigma,\sigma)\le\eps)\ge\tfrac12$, then, with $w_d(t)=t\sqrt{\min\{t,d^2\}}$,
\[
\sup_{\sigma\preceq2I/d}\ \E_\sigma\Big[\sum_{j=1}^Tw_d(t_j)\Big]=\Omega\Big(\frac{d^3}{\eps^2}\Big).
\]
The proof in \cite{CMMP26} uses a prior supported on $\{\sigma=I/d+\Delta:\|\Delta\|_\op\le C_{\rm fd}\,\eps/d\}$ and nowhere else the promise $\sigma\preceq2I/d$, so the conclusion also holds when the success requirement is imposed only on that operator-norm ball; we refer to this as Lemma~\ref{lem:fd}$'$.
\end{lemma}

\section{At the maximally mixed state: a replicated-parameter Fisher budget}\label{sec:mixed}

Theorem~1.1 of \cite{Rub26} rests on Proposition~3.4 there: for every POVM $M$ on one copy and every $\|\theta\|_2\le r$, the classical Fisher information matrix $I_M(\theta)$ of the outcome law of $M$ on $\rho_\theta$ satisfies $\tr I_M(\theta)\le C_r^2m$, with $C_r=\sinh(2r)/(2r)$ and $C_r^2<2$ for $r\le r_0=1/(10\sqrt3)$. We extend this to joint measurements on $t$ copies.

\begin{lemma}[$t$ copies are one Gaussian state on $mt$ modes]\label{lem:replicate}
Fix $t\ge1$. On $(\C^2)^{\otimes mt}$, with Jordan--Wigner Majoranas $c_1,\dots,c_{2mt}$ as in \cite[Sec.~2]{Rub26}, index the modes by pairs $(j,c)$, $1\le j\le m$, $1\le c\le t$, so that the $2m$ Majoranas of block $c$ are $c^{(c)}_a=P^{\otimes(c-1)}\otimes c_a\otimes I^{\otimes(t-c)}$ with $P$ the parity operator of $m$ modes. Then $B^{(c)}_{ab}:=ic^{(c)}_ac^{(c)}_b=I^{\otimes(c-1)}\otimes B_{ab}\otimes I^{\otimes(t-c)}$, and for $\theta\in\R^p$ the $mt$-mode parameter $\Theta=A\theta\in\R^P$, $P=mt(2mt-1)$, defined by $\Theta_{(a,c),(b,c)}=\theta_{ab}$ for each $c$ and $\Theta=0$ on all cross-block pairs, satisfies
\[
\rho_\Theta=\rho_\theta^{\otimes t},\qquad A^TA=tI_p,\qquad\|AA^T\|_\op=t .
\]
The normal-form values of $\Theta$ are those of $\theta$, each with multiplicity $t$.
\end{lemma}

\begin{proof}
$P^2=I$ gives the product formula for $B^{(c)}_{ab}$, and these operators on different blocks commute. Hence $K_\Theta=\sum_cK^{(c)}_\theta$ is a sum of commuting terms and $e^{K_\Theta}=\bigotimes_ce^{K_\theta}$; the normalised trace is multiplicative, $z_\Theta=z_\theta^t$, and $2^{mt}=(2^m)^t$, so $\rho_\Theta=e^{K_\Theta}/(2^{mt}z_\Theta)=\rho_\theta^{\otimes t}$. $A$ is a $P\times p$ matrix with $t$ entries equal to $1$ in each column and at most one in each row, so $A^TA=tI_p$ and $AA^T$ is $t$ times a projector. Applying the rotation of \cite[Lemma~2.1]{Rub26} within each block gives $K_\Theta=\sum_c\sum_j\lambda_jZ^{(c)}_j$ with $mt$ commuting involutions, which is the normal form of $\Theta$.
\end{proof}

\begin{lemma}[block Fisher budget]\label{lem:blockfisher}
Let $t\ge1$, $\|\theta\|_2\le r$, and let $M$ be any POVM on $t$ copies together with ancillas in a $\theta$-independent state. Let $I^{(t)}_M(\theta)$ be the Fisher information matrix, with respect to $\theta\in\R^p$, of the outcome law of $M$ on $\rho_\theta^{\otimes t}$. Then
\[
\tr I^{(t)}_M(\theta)\le C_r^2\,m\,t^2,
\]
and in particular $\tr I^{(t)}_M(\theta)\le2mt^2$ for $\|\theta\|_2\le r_0$.
\end{lemma}

\begin{proof}
By Lemma~\ref{lem:replicate}, $\rho_\theta^{\otimes t}=\rho_{A\theta}$ in the $mt$-mode family, so by the chain rule $\partial_{\theta_j}\rho_\theta^{\otimes t}=\sum_J A_{Jj}\partial_{\Theta_J}\rho_\Theta$ and $I^{(t)}_M(\theta)=A^TI^{(mt)}_M(\Theta)A$, where $I^{(mt)}_M(\Theta)$ is the Fisher matrix of the $mt$-mode family at $\Theta$. Since $I^{(mt)}_M\succeq0$, $\tr(A^TI^{(mt)}_MA)=\tr(I^{(mt)}_MAA^T)\le\|AA^T\|_\op\tr I^{(mt)}_M(\Theta)=t\,\tr I^{(mt)}_M(\Theta)$. It remains to bound $\tr I^{(mt)}_M(\Theta)$ by $C_r^2mt$. Proposition~3.4 of \cite{Rub26}, applied to the $mt$-mode family, gives $\tr I^{(mt)}_M(\Theta)\le\|\Tcal_\Theta|_V\|^2\cdot mt$ by its proof, where $V$ is the span of the $mt$-mode quadratics, $\Tcal_\Theta=g(\ad K_\Theta)$ with $g(x)=\sinh(x/2)/(x/2)$, and the factor $mt$ is the covariance bound \cite[Lemma~3.1]{Rub26} on $mt$ modes; and \cite[Lemma~3.3]{Rub26} gives $\|\ad K_\Theta|_V\|=2\max_{J<K}(\lambda_J+\lambda_K)$ over the normal-form values of $\Theta$ when $mt\ge2$; for $mt=1$ the quadratic algebra is abelian, the commutator norm is $0$, and the bound below holds trivially. By Lemma~\ref{lem:replicate} these are the $\lambda_j(\theta)$ with multiplicity $t$, so $\max_{J<K}(\lambda_J+\lambda_K)\le2\max_j\lambda_j\le2\|\theta\|_2\le2r$ (for $t\ge2$ the multiplicity makes $J<K$ with $\lambda_J=\lambda_K=\lambda_{\max}$ available, which is why the bound is $2\lambda_{\max}$ rather than the $\sqrt2\|\theta\|_2$ of the single-copy case; for $t=1$ the bound $2\lambda_{\max}$ still holds, less sharply). Hence $\|\ad K_\Theta|_V\|\le4r$ and, $g$ being even and increasing in $|x|$, $\|\Tcal_\Theta|_V\|\le g(4r)=C_r$ exactly as in \cite[Eq.~(3.2)]{Rub26}. Note that the hypothesis is on $\|\theta\|_2$, not on $\|\Theta\|_2=\sqrt t\|\theta\|_2$; this is the one place where the proof of \cite[Prop.~3.4]{Rub26} is re-read rather than its statement invoked. For $r\le r_0$, $C_r^2<2$ as there.
\end{proof}

\begin{remark}
The bound $2mt^2$ cannot be the whole story for large $t$: the Fisher trace of any measurement is at most the trace of the quantum Fisher information, which for $\rho_\theta^{\otimes t}$ is at most $tp=tm(2m-1)$ by Lemma~\ref{lem:qfi} below, so for $t\ge m$ the quantum bound is the smaller. Whether a joint measurement on $t<m$ copies can attain Fisher trace of order $mt^2$ is answered at $\theta=0$ by Theorem~\ref{thm:exact} below: for $t=2$ the budget is attained as $m\to\infty$, and the exact values for $t\le6$ approach it. On the ball around $\theta=0$ the question is open; its relation to whether the $1/k$ of Theorem~\ref{thm:kcopy-mixed} can be improved to the $1/\sqrt k$ that holds near the vacuum (Theorem~\ref{thm:vacuum} below) is discussed in Remark~\ref{rem:tight}.
\end{remark}

\begin{theorem}[fermions at the maximally mixed state, $k$ copies]\label{thm:kcopy-mixed}
Let $0<\eps\le1/3600$, $m\ge1$ and $k\ge1$. Consider a $k$-copy protocol that receives $N$ copies of an unknown mixed fermionic Gaussian state $\rho$ on $m$ modes and outputs a density matrix $\hat\rho$, not necessarily Gaussian, with $T(\hat\rho,\rho)\le\eps$ with probability at least $2/3$ for every such $\rho$. Then
\[
N>\frac{p^2}{8{,}553{,}600\,m\,k\,\eps^2}\ \ge\ \frac{m^3}{8{,}553{,}600\,k\,\eps^2},\qquad p=m(2m-1).
\]
The hard family lies in $\{\rho_\theta:\|\theta\|_2\le120\sqrt3\,\eps\}$, so the bound holds on every class containing it, in particular on $\Gc$ for every $c\le1-\tanh(120\sqrt3\,\eps)$ and on the Gibbs classes $\mathcal C(\beta_0)$ of Section~\ref{sec:nowarm} below for $\beta_0\ge120\sqrt3\,\eps$.
\end{theorem}

\begin{proof}
The proof of \cite[Thm.~1.1]{Rub26} is followed step by step, with blocks in place of copies.

\emph{Adaptivity.} Let the protocol use $n$ copies in blocks $t_1,t_2,\dots$ with $t_j\le k$ and $\sum_jt_j\le n$; pad a protocol that stops early with trivial blocks of size $1$ and POVM $\{I\}$, which contribute nothing to the score, until exactly $n$ copies are consumed. The number of blocks is then random, but the event that block $j$ exists is determined by the history $h_j$, so the chain rule below is unaffected. Conditional on the classical history $h_j$ before block $j$, the block is a fresh system in the state $\rho_\theta^{\otimes t_j}\otimes\eta_j$ with $\eta_j$ independent of $\theta$, unentangled with everything measured so far. The chain-rule argument of \cite[Lemma~4.1]{Rub26} applies verbatim with ``copy'' read as ``block'': the transcript score is the sum of the block scores, each block score has conditional mean zero, cross terms vanish by the tower property, and
\[
\tr I_Y(\theta)=\sum_j\E_{h_j}\big[\tr I^{(t_j)}_{M^{(h_j)}}(\theta)\big]\le\sum_j\E_{h_j}\big[2mt_j^2\big]\le2mk\,\E\Big[\sum_jt_j\Big]\le2mkn
\]
for $\|\theta\|_2\le r_0$, by Lemma~\ref{lem:blockfisher} and $t_j\le k$. The two points of rigour in that proof, bounded scores and dominated convergence, hold for the same reasons on the $2^{mt_j}$-dimensional block space.

\emph{Van Trees.} With the prior of \cite[Lemma~6.1]{Rub26} on the cube $[-a,a]^p$, $a=r/\sqrt p$, $r\le r_0$, the vector van Trees inequality \cite[Lemma~6.2]{Rub26} reads, with the budget $2mkn$ in place of $2mn$,
\[
\E\|\hat\theta-\theta\|_2^2\ge\frac{p^2}{2mkn+\pi^2p^2/r^2},\qquad\text{so }\ \E\|\hat\theta-\theta\|_2^2>0.09\,r^2\ \text{ whenever }\ n\le\frac{p^2}{2mkr^2} .
\]

\emph{Amplification and projection.} \cite[Lemma~6.3]{Rub26} applies unchanged to a $k$-copy protocol, since the $99$ runs use fresh copies and only classical data passes between them: for $\theta$ in the cube, the projected estimator $\hat\theta$ satisfies $\|\hat\theta-\theta\|_2\le6\eps/\kappa$ with probability at least $0.99$, and $\le2r$ always, with $\kappa=1/(4\sqrt3)$.

\emph{Conclusion.} Set $r=120\sqrt3\,\eps$, so $r\le r_0$ is exactly $\eps\le1/3600$, and $6\eps/\kappa=r/5$. The amplified protocol is a $k$-copy protocol with $n=99N$ copies. Its risk is at most $(r/5)^2+0.01(2r)^2=0.08\,r^2$ by the amplification lemma, and at least $0.09\,r^2$ by the van Trees step if $99N\le p^2/(2mkr^2)$. Hence $99N>p^2/(2mkr^2)$, that is $N>p^2/(198\cdot43{,}200\,mk\eps^2)=p^2/(8{,}553{,}600\,mk\eps^2)$, and $p\ge m^2$ gives the second form. Every state in play is $\rho_\theta$ with $\|\theta\|_2\le r$, which gives the statement about the hard family; its normal-form values are at most $\tanh r$.
\end{proof}

\begin{remark}
For $k=1$ this is \cite[Thm.~1.1]{Rub26}. For $k\ge m$ the bound falls below the collective lower bound of Theorem~\ref{thm:collective} below, which holds for every protocol; so Theorem~\ref{thm:kcopy-mixed} is informative for $1\le k<m$, whereas the near-vacuum bound of Theorem~\ref{thm:vacuum} below is informative up to $k<m^2$. On $\Gc$ for fixed $c$ the $k$-copy rate therefore lies in $[\Omega(\max\{m^3/k,m^2\}),O(m^3)]/\eps^2$.
\end{remark}

\subsection{The exact block budget at the maximally mixed state}\label{sec:exact}

Whether the $1/k$ of Theorem~\ref{thm:kcopy-mixed} can be improved by this method depends on whether the budget $2mt^2$ of Lemma~\ref{lem:blockfisher} is loose. At $t=2$ it is not, up to a constant factor, and the exact values for $t\le6$ approach $mt^2$ as $m$ grows. At $\theta=0$ the supremum of the Fisher trace over all measurements on $t$ copies can be computed exactly in terms of the representation theory of $\mathfrak{so}(2m)$, and the answer is the dimension-weighted average of the squared highest weight.

Let $\mathfrak{so}(2m)$ act on the Fock space $\Fock_m\cong\C^{2^m}$ through the quadratics $B_{ab}$; this is the Dirac spinor representation $S$, the sum of the two half-spin representations, and the Gaussian unitaries are the group it generates. On $t$ copies the operators $G_{ab}=\sum_cB^{(c)}_{ab}$ of Lemma~\ref{lem:replicate} give the diagonal action on $S^{\otimes t}$, which decomposes into irreducible representations $V_\lambda$ with multiplicities $m_\lambda$. We use the normalisation in which the Cartan generators are $H_j=\sum_cZ^{(c)}_j$, so that the weights of $S$ are the vectors in $\{\pm1\}^m$ and a highest weight $\lambda\in\mathbb Z^m$ satisfies $\lambda_1\ge\dots\ge\lambda_{m-1}\ge|\lambda_m|$; we write $\|\lambda\|^2=\sum_j\lambda_j^2$.

\begin{theorem}[exact block budget at $\theta=0$]\label{thm:exact}
For $m\ge1$ and $t\ge1$,
\[
F_{\max}(m,t):=\sup_M\tr I^{(t)}_M(0)=\frac1{2^{mt}}\sum_\lambda m_\lambda\dim V_\lambda\,\|\lambda\|^2 ,
\]
the supremum over all POVMs on $t$ copies with ancillas, and the sum over the decomposition $S^{\otimes t}=\bigoplus_\lambda m_\lambda V_\lambda$ under $\mathfrak{so}(2m)$. The supremum is attained by the measurement that, in each irreducible summand, is the Haar-covariant family of highest-weight (coherent) states.
\end{theorem}

\begin{proof}
At $\theta=0$, $\rho_0^{\otimes t}=I/2^{mt}$, $\partial_{ab}\rho_\theta^{\otimes t}|_{0}=G_{ab}/2^{mt}$, and for a POVM $\{M_x\}$ with $q(x)=\Tr M_x/2^{mt}$ and $\sigma_x=M_x/\Tr M_x$ the score is $\Tr(\sigma_xG_{ab})$, so $\tr I^{(t)}_M(0)=\sum_xq(x)\,\|\mu(\sigma_x)\|^2$ with $\mu(\sigma)_{ab}=\Tr(\sigma G_{ab})$, the moment map; ancillas are absorbed as in \cite[Sec.~3]{Rub26}.

\emph{Upper bound.} Let $\Pi_\lambda$ be the projector onto the $\lambda$-isotypic component. Each $G_{ab}$ commutes with every $\Pi_\lambda$, so $\mu(\sigma)=\sum_\lambda p_\lambda\mu(\sigma_\lambda)$ with $p_\lambda=\Tr(\Pi_\lambda\sigma)$ and $\sigma_\lambda=\Pi_\lambda\sigma\Pi_\lambda/p_\lambda$, and by convexity of $\|\cdot\|^2$, $\|\mu(\sigma)\|^2\le\sum_\lambda p_\lambda\|\mu(\sigma_\lambda)\|^2$. A state on the isotypic component $V_\lambda\otimes\C^{m_\lambda}$ has the same moment map as its reduction to $V_\lambda$, which is a convex combination of pure states of $V_\lambda$, so it suffices to show $\|\mu(\psi)\|^2\le\|\lambda\|^2$ for unit $\psi\in V_\lambda$. The matrix $X_{ab}=\langle\psi|G_{ab}|\psi\rangle$ is real antisymmetric, and the action is covariant: for a Gaussian unitary $U_O$, $O\in SO(2m)$, $\mu(U_O\psi)=O\mu(\psi)O^T$ while $U_O\psi\in V_\lambda$ still. Choose $O\in SO(2m)$ bringing $X$ to the normal form $\bigoplus_jx_jJ$ (the normal form is reachable in $SO(2m)$, since flipping one basis vector of a block changes only the sign of one $x_j$); $U_O$ is in the group generated by the $G_{ab}$, so it preserves each irreducible summand, and no separate treatment of the two half-spin components is needed. Then $\|\mu(\psi)\|^2=\sum_jx_j^2$ with $x_j=\langle\psi|H_j|\psi\rangle$. Expanding $\psi$ in an orthonormal basis of weight vectors, $x_j=\sum_\nu|c_\nu|^2\nu_j$, so $(x_j)_j$ lies in the convex hull of the weights of $V_\lambda$, hence in the convex hull of the Weyl orbit of $\lambda$ \cite[Sec.~13.4]{Hum72}; the Euclidean norm is convex and the Weyl group acts orthogonally, so $\sum_jx_j^2\le\|\lambda\|^2$. Therefore $\tr I^{(t)}_M(0)\le\sum_xq(x)\sum_\lambda\Tr(\Pi_\lambda\sigma_x)\|\lambda\|^2=\sum_\lambda\|\lambda\|^2\Tr(\Pi_\lambda\,\textstyle\sum_xq(x)\sigma_x)=\sum_\lambda\|\lambda\|^2\dim(\Pi_\lambda)/2^{mt}$, since $\sum_xq(x)\sigma_x=I/2^{mt}$ and $\dim\Pi_\lambda=m_\lambda\dim V_\lambda$.

\emph{Attainment.} Fix a decomposition into irreducible subspaces $V_{\lambda,i}$, $i\le m_\lambda$, with highest-weight unit vectors $v_{\lambda,i}$, and let $g\mapsto U_g$ be the spin representation of the compact group generated by the $G_{ab}$. By Schur's lemma $\int U_g|v_{\lambda,i}\rangle\langle v_{\lambda,i}|U_g^\dagger\,dg=I_{V_{\lambda,i}}/\dim V_\lambda$ for Haar measure, so $\{\dim V_\lambda\,U_g|v_{\lambda,i}\rangle\langle v_{\lambda,i}|U_g^\dagger\,dg\}_{\lambda,i,g}$ is a POVM on $S^{\otimes t}$ with outcome density $q(\lambda,i,g)=\dim V_\lambda/2^{mt}$. Its post-measurement states are $U_gv_{\lambda,i}$, with $\|\mu(U_gv_{\lambda,i})\|^2=\|\mu(v_{\lambda,i})\|^2=\|\lambda\|^2$, because $\langle v_\lambda|G|v_\lambda\rangle$ vanishes on root vectors and equals $\lambda_j$ on $H_j$. Summing gives the right-hand side.
\end{proof}

\begin{corollary}[two copies]\label{cor:twocopies}
$F_{\max}(m,2)=4m\big(1-\binom{2m}{m}4^{-m}\big)$. In particular $F_{\max}(m,2)\ge3m$ for $m\ge5$ and $F_{\max}(m,2)/(m\cdot2^2)\to1$ as $m\to\infty$: for blocks of two copies the budget $mt^2$ of Lemma~\ref{lem:blockfisher} is attained asymptotically.
\end{corollary}

\begin{proof}
$S\otimes S\cong\operatorname{End}(S)\cong\mathrm{Cl}(2m)\cong\Lambda(\C^{2m})=\bigoplus_{k=0}^{2m}\Lambda^k(\C^{2m})$ as $\mathfrak{so}(2m)$-modules, $S$ being self-dual. For $k\le m$ the highest weight of $\Lambda^k$ is $(2,\dots,2,0,\dots,0)$ with $k$ twos, of squared norm $4k$ (for $k=m$ the two summands have highest weights $(2,\dots,2,\pm2)$, both of squared norm $4m$), and $\Lambda^{2m-k}\cong\Lambda^k$. Hence $F_{\max}(m,2)=4^{-m}\sum_{k=0}^{2m}\binom{2m}{k}\,4\min\{k,2m-k\}=4^{1-m}\big(m4^m-\sum_k\binom{2m}{k}|k-m|\big)$, and $\sum_k\binom{2m}{k}|k-m|=m\binom{2m}{m}$ by the identity $\sum_{k>m}k\binom{2m}{k}=m4^m/2$. Stirling gives $\binom{2m}{m}4^{-m}\sim(\pi m)^{-1/2}$, and $\binom{10}{5}4^{-5}=0.246$.
\end{proof}

The same conclusion holds for every fixed block size, with an explicit error term. The proof uses only the quadratic Casimir operator and the fact that highest weights of $S^{\otimes t}$ have entries at most $t/2$.

\begin{theorem}[the budget is asymptotically tight for every fixed block size]\label{thm:asymptotic}
For all $m\ge1$ and $t\ge1$,
\[
mt^2-\sqrt2\,t^2\sqrt{m(t-1)}-3t^2\ \le\ F_{\max}(m,t)\ \le\ mt^2 .
\]
In particular $F_{\max}(m,t)/(mt^2)\to1$ as $m\to\infty$ for every fixed $t$, and $F_{\max}(m,t)\ge\tfrac12mt^2$ whenever $m\ge32t$.
\end{theorem}

\begin{proof}
The upper bound is Lemma~\ref{lem:blockfisher} at $\theta=0$, where $C_0=1$. For the lower bound write $\lambda_s=\lambda/2$ for the highest weights in the standard normalisation, in which the weights of $S$ are $(\pm\tfrac12,\dots,\pm\tfrac12)$, the positive roots of $\mathfrak{so}(2m)$ are $e_i\pm e_j$ ($i<j$) and half their sum is $\delta=(m-1,m-2,\dots,0)$. By Theorem~\ref{thm:exact}, $F_{\max}(m,t)=4\,\E\|\lambda_s\|^2$, where $\E$ is the expectation under the probability $P(\lambda)=m_\lambda\dim V_\lambda/2^{mt}$ on the highest weights of $S^{\otimes t}$.

\emph{Casimir.} For any nondegenerate invariant bilinear form on $\mathfrak{so}(2m)$, the associated Casimir element acts on $V_\lambda$ as the scalar $\langle\lambda_s+\delta,\lambda_s+\delta\rangle-\langle\delta,\delta\rangle=\langle\lambda_s,\lambda_s+2\delta\rangle$, the inner product on $\mathfrak h^*$ being the one dual to the form \cite[Sec.~10.2]{Hal15}, \cite[Thm.~5.5 and Prop.~5.28]{Kna02}. Take the form in which the generators $B_{ab}/2$ are orthonormal; its Casimir is $C_2=\sum_{a<b}(G_{ab}/2)^2$ on $S^{\otimes t}$, and the dual inner product is the one in which the $e_j$ are orthonormal, as the check on $S$ confirms: there $C_2=\sum_{a<b}(B_{ab}/2)^2=p/4=\langle\omega,\omega+2\delta\rangle$ for $\omega=(\tfrac12,\dots,\tfrac12)$, and on $S\otimes S=\Lambda(\C^{2m})$ one finds $C_2(\Lambda^k)=k(2m-k)$ with binomial average $m^2-m/2=p/2$, as required below at $t=2$. (For $m=1$ the statement is trivial, and for $m=2$ both simple factors of $\mathfrak{so}(4)$ are seen by $S=S_+\oplus S_-$.) On the other hand $G_{ab}^2=tI+\sum_{c\ne c'}B^{(c)}_{ab}B^{(c')}_{ab}$ and the cross terms are traceless, so $\Tr_{S^{\otimes t}}C_2=2^{mt}\,tp/4$. Dividing by $2^{mt}$,
\begin{equation}\label{eq:casimir}
\E\|\lambda_s\|^2+2\,\E\langle\lambda_s,\delta\rangle=\frac{tp}{4}.
\end{equation}

\emph{Deficits.} Every highest weight of $S^{\otimes t}$ satisfies $\tfrac t2\ge\lambda_{s,1}\ge\dots\ge\lambda_{s,m-1}\ge|\lambda_{s,m}|$, since all weights of $S^{\otimes t}$ lie in $[-\tfrac t2,\tfrac t2]^m$. Put $d_j=\tfrac t2-\lambda_{s,j}\in[0,\tfrac t2]$ for $j<m$, $S'=\sum_{j<m}d_j$ and $\Delta=\sum_{j<m}d_j(m-j)\ge0$. Then
\[
\frac{mt^2}{4}-\|\lambda_s\|^2=\sum_{j<m}d_j(t-d_j)+\Big(\frac{t^2}{4}-\lambda_{s,m}^2\Big)\le tS'+\frac{t^2}{4},\qquad
\frac t2\sum_{j=1}^m(m-j)-\langle\lambda_s,\delta\rangle=\Delta ,
\]
the second because the $j=m$ term of $\delta$ vanishes. Substituting both identities into \eqref{eq:casimir} and using $\tfrac{mt^2}4+\tfrac{tm(m-1)}2-\tfrac{tp}4=\tfrac{mt(t-1)}4$,
\[
\E\Big[\sum_{j<m}d_j(t-d_j)+\frac{t^2}{4}-\lambda_{s,m}^2\Big]+2\,\E\Delta=\frac{mt(t-1)}{4}.
\]
Both expectations are of nonnegative quantities, so $\E\Delta\le mt(t-1)/8$.

\emph{$\Delta$ controls $S'$.} For fixed $S'$, the sum $\sum_{j<m}d_j(m-j)$ over $d\in[0,\tfrac t2]^{m-1}$ with $\sum_jd_j=S'$ is minimised by filling the positions of smallest weight first, $j=m-1,m-2,\dots$: with $L=\lfloor2S'/t\rfloor$ full entries, $\Delta\ge\tfrac t2(1+2+\dots+L)\ge tL^2/4\ge\tfrac t4(2S'/t-1)^2$ when $2S'\ge t$. Hence $S'\le\sqrt{t\Delta}+t/2$ in every case.

\emph{Conclusion.} By the two preceding steps and Jensen's inequality for the concave square root,
\[
\E\Big[\frac{mt^2}{4}-\|\lambda_s\|^2\Big]\le t\,\E S'+\frac{t^2}{4}\le t^{3/2}\sqrt{\E\Delta}+\frac{3t^2}{4}\le t^{3/2}\sqrt{\frac{mt(t-1)}{8}}+\frac{3t^2}{4},
\]
and multiplying by $4$ gives $mt^2-F_{\max}(m,t)\le\sqrt2\,t^2\sqrt{m(t-1)}+3t^2$. For $m\ge32t$ the right-hand side is at most $mt^2(\tfrac14+\tfrac3{32})<\tfrac12mt^2$.
\end{proof}

The bound is far from the exact values at small $m$ (Table~\ref{tab:fmax}); its content is the limit. In the proof, the Casimir identity \eqref{eq:casimir} says that the typical highest weight falls short of the top weight $(t,\dots,t)$ by an amount that, weighted by $\delta$ and measured in the standard normalisation, is at most $mt(t-1)/8$ on average; since $\delta_j=m-j$ is small only near $j=m$, the shortfall must sit in the last $O(\sqrt{mt})$ coordinates, and there it costs only $O(t^2\sqrt{mt})$ in squared norm.

\begin{table}[t]
\centering\small
\begin{tabular}{@{}r|rrrrrr@{}}
\toprule
$m\backslash t$ & 1 & 2 & 3 & 4 & 5 & 6\\
\midrule
2 & 1 & 2.500 & 4.250 & 6.156 & 8.172 & 10.270\\
4 & 1 & 2.906 & 5.352 & 8.221 & 11.415 & 14.871\\
6 & 1 & 3.098 & 5.907 & 9.319 & 13.208 & 17.506\\
8 & 1 & 3.214 & 6.260 & 10.034 & 14.400 & 19.291\\
10 & 1 & 3.295 & 6.510 & 10.551 & 15.273 & 20.612\\
\midrule
$t^2$ & 1 & 4 & 9 & 16 & 25 & 36\\
$t^{3/2}$ & 1 & 2.828 & 5.196 & 8 & 11.180 & 14.697\\
\bottomrule
\end{tabular}
\caption{$F_{\max}(m,t)/m$ from Theorem~\ref{thm:exact}, computed exactly from the decomposition of $S^{\otimes t}$ (multiplicities by the Weyl alternating formula in determinant form, dimensions by the Weyl dimension formula; the values are rationals, e.g.\ $F_{\max}(2,2)=5$, $F_{\max}(3,2)=33/4$, $F_{\max}(2,3)=17/2$). For fixed $t$ the ratio increases with $m$ (for $t=3$ it is $0.63,0.72,0.80,0.85$ times $t^2$ at $m=5,10,20,40$); for fixed $m$ it appears to grow like $t^{3/2}$ once $t$ is comparable with $m$, and it must eventually fall below $tp/m=t(2m-1)$, the quantum Fisher bound of Lemma~\ref{lem:qfi}.}
\label{tab:fmax}
\end{table}

\begin{remark}[what this says about Theorem~\ref{thm:kcopy-mixed}]\label{rem:tight}
Theorems~\ref{thm:exact} and~\ref{thm:asymptotic} show that no argument that bounds the Fisher information of a block of $t$ copies at the maximally mixed state can improve on the budget $2mt^2$ of Lemma~\ref{lem:blockfisher} by more than the factor $4$ for $m\ge32t$, and Corollary~\ref{cor:twocopies} gives the factor $8/3$ at $t=2$, $m\ge5$; Table~\ref{tab:fmax} shows the ratio $F_{\max}(m,t)/(mt^2)$ increasing with $m$ for every $t\le6$, reaching $0.85$ at $t=3$, $m=40$. Since $F_{\max}(m,t)/(mt^{3/2})\ge\tfrac12\sqrt t$ for $m\ge32t$, the uniform scalar Fisher-trace budget on which Theorem~\ref{thm:kcopy-mixed} rests is sharp up to the factor $4$ for $t\le m/32$: a bound $\tr I^{(t)}_M\le Cmt^{3/2}$ holding for all measurements on a neighbourhood of $\theta=0$ fails already at $\theta=0$, and, the Fisher information of the covariant measurement of Theorem~\ref{thm:exact} being continuous in $\theta$ (its outcome density is strictly positive), on a neighbourhood whose size the argument does not control. So the $1/k$ of Theorem~\ref{thm:kcopy-mixed} is the best that the budget route can give at $I/2^m$; a lower bound with the dependence $1/\sqrt k$ would need information beyond a pointwise trace budget, such as a prior-averaged or matrix-valued inequality. This is the opposite of the situation for unstructured $d$-dimensional states, where the block Fisher information at the maximally mixed state is at most of order $d^2t^{3/2}$ \cite[Thm.~8.1]{KLMR26}, the right order given the matching upper bounds, and of the near-vacuum Gaussian class, whose rate is $m^3/\sqrt k$ by Theorem~\ref{thm:vacuum} and Theorem~\ref{cor:kcopy-rate} below. The reason is visible in the proof: a single Gaussian copy has Fisher budget $m$ against a quantum Fisher information $p\approx2m^2$, so there is room for a factor $t$ of coherent gain per copy before the quantum bound binds at $t\approx m$, whereas for unstructured states the gap between $d$ and $d^2$ is filled only as $\sqrt t$. The author conjectures that the $k$-copy rate at the maximally mixed state is $\Theta(m^3/(k\eps^2))$ for $k\le m$, strictly faster than near the vacuum; the lower bound is Theorem~\ref{thm:kcopy-mixed}, and what is missing is a $k$-copy protocol whose estimator attains the covariant Fisher information of Theorem~\ref{thm:exact}, together with an extension of Theorems~\ref{thm:exact} and~\ref{thm:asymptotic} from $\theta=0$ to the ball $\|\theta\|_2\le r_0$.
\end{remark}

\subsection{The collective rate at the maximally mixed state}

The single-copy budget $2m$ of \cite[Prop.~3.4]{Rub26} has a collective counterpart: the quantum Fisher information of the family \eqref{eq:family}, which bounds the Fisher information of every measurement whatever its structure. Its trace is at most $p=m(2m-1)$, so the factor $m$ between the single-copy and collective rates is the ratio of the two budgets. The collective lower bound $\Omega(m^2/\eps^2)$ of \cite[Thm.~4.3]{CFGLMWW26} is proved on pure states; the following gives it at the maximally mixed state, with explicit constants, on every class containing the hard family of Theorem~\ref{thm:kcopy-mixed}.

\begin{lemma}[quantum Fisher budget]\label{lem:qfi}
For every $\theta\in\R^p$ the symmetric-logarithmic-derivative quantum Fisher information matrix $F_Q(\theta)$ of the family \eqref{eq:family} satisfies $\tr F_Q(\theta)\le p$, with equality at $\theta=0$.
\end{lemma}

\begin{proof}
The SLD metric is the smallest of the monotone metrics and the Bogoliubov--Kubo--Mori metric is one of them \cite{Pet96,LR99}, so $F_Q(\theta)\preceq F_{\rm BKM}(\theta)$, where
\[
F_{\rm BKM}(\theta)_{jl}=\int_0^1\Tr\big(\rho_\theta^s\,L_j\,\rho_\theta^{1-s}\,L_l\big)\,ds,\qquad L_j=\partial_j\log\rho_\theta=B_j-\mu_jI,
\]
the last identity because $\log\rho_\theta=K_\theta-\log(dz_\theta)$ by \eqref{eq:family}, with $\mu_j=\Tr(\rho_\theta B_j)$. For a density matrix $\rho=\sum_ip_i|i\rangle\langle i|$ and Hermitian $X$, $\Tr(\rho^sX\rho^{1-s}X)=\sum_{i,i'}p_i^sp_{i'}^{1-s}|X_{ii'}|^2\le\sum_{i,i'}(sp_i+(1-s)p_{i'})|X_{ii'}|^2=\Tr(\rho X^2)$ by the weighted arithmetic--geometric mean inequality and $\sum_{i'}|X_{ii'}|^2=(X^2)_{ii}$. Hence $\tr F_{\rm BKM}(\theta)\le\sum_j\Tr(\rho_\theta(B_j-\mu_j)^2)=\sum_j(1-\mu_j^2)\le p$, using $B_j^2=I$. At $\theta=0$, $\rho_0=I/d$, the SLDs are the orthonormal $B_j$ and $\tr F_Q(0)=p$.
\end{proof}

\begin{theorem}[collective lower bound, explicit constants]\label{thm:collective}
Let $0<\eps\le1/3600$ and $m\ge1$. Any protocol whatsoever, with joint measurements on all copies, quantum memory, ancillas and adaptivity allowed, that learns every mixed fermionic Gaussian state on $m$ modes to trace distance $\eps$ with probability at least $2/3$ from $N$ copies satisfies
\[
N>\frac{p}{4{,}276{,}800\,\eps^2}\ \ge\ \frac{m^2}{4{,}276{,}800\,\eps^2}.
\]
The hard family is that of Theorem~\ref{thm:kcopy-mixed}, $\{\rho_\theta:\|\theta\|_2\le120\sqrt3\,\eps\}$, so the bound holds on every class containing it.
\end{theorem}

\begin{proof}
After fixing its internal randomness, which has a $\theta$-independent law and contributes nothing to the score, a protocol with $n$ copies is one POVM on $\rho_\theta^{\otimes n}\otimes\eta$ with $\eta$ independent of $\theta$. By the Braunstein--Caves inequality \cite{BC94} the Fisher matrix of its outcome law is at most $F_Q(\rho_\theta^{\otimes n}\otimes\eta)=nF_Q(\theta)$, the quantum Fisher information being additive on products and zero on the $\theta$-independent factor; so $\tr I_Y(\theta)\le np$ for every $\theta$ by Lemma~\ref{lem:qfi}; the scores are bounded and dominated convergence applies as in \cite[Lemma~4.1]{Rub26}, since $\rho_\theta^{\otimes n}$ has full rank. The van Trees inequality \cite[Lemma~6.2]{Rub26} with the budget $np$ in place of $2mn$ gives $\E\|\hat\theta-\theta\|_2^2\ge p^2/(np+\pi^2p^2/r^2)\ge r^2/(1+\pi^2)>0.09\,r^2$ whenever $n\le p/r^2$. The amplification lemma \cite[Lemma~6.3]{Rub26} applies to any protocol, since the $99$ runs use fresh copies and exchange only classical data, and gives risk at most $0.08\,r^2$ with $r=120\sqrt3\,\eps$ as in the proof of Theorem~\ref{thm:kcopy-mixed}. Hence $99N>p/r^2=p/(43{,}200\,\eps^2)$, and $99\cdot43{,}200=4{,}276{,}800$.
\end{proof}

With Theorem~\ref{thm:collective} the class $\Gc$ for fixed $c$ is characterised at both ends: $\Theta(m^2/\eps^2)$ for collective protocols, by \cite[Thm.~4.2]{CFGLMWW26} and Theorem~\ref{thm:collective}, and $\Theta(m^3/\eps^2)$ for single-copy protocols, by \cite[Thms.~1.1 and~7.6]{Rub26}; the two Fisher budgets $p$ and $2m$ differ by the factor $m$ that separates the rates.

\section{No warm regime}\label{sec:nowarm}

For $\beta_0>0$ let
\[
\mathcal C(\beta_0)=\{\rho_\theta:\|\theta\|_2\le\beta_0\}=\Big\{\frac{e^{-\beta H}}{\Tr e^{-\beta H}}:\ H=\sum_{a<b}h_{ab}B_{ab},\ \beta\|h\|_F\le\beta_0\Big\},
\]
the quadratic Gibbs states whose inverse temperature, in units of the Frobenius norm of the coefficient matrix, is at most $\beta_0$. As $\beta_0\to0$ every state of $\mathcal C(\beta_0)$ tends to $I/2^m$, the infinite-temperature state of every quadratic Hamiltonian.

\begin{theorem}[no warm regime]\label{prop:nowarm}
Let $\beta_0>0$ and $k\ge1$.
\begin{enumerate}
\item[(i)] For $0<\eps\le\min\{1/3600,\ \beta_0/(120\sqrt3)\}$, every $k$-copy protocol learning $\mathcal C(\beta_0)$ to trace distance $\eps$ with probability $2/3$ uses $N>m^3/(8{,}553{,}600\,k\,\eps^2)$ copies; the constant does not depend on $\beta_0$.
\item[(ii)] $\mathcal C(\beta_0)\subset\mathcal G_{1-\tanh\beta_0}$, so single-copy matchgate shadows learn it with $O(m^3/((1-\tanh\beta_0)\eps^2))$ copies \cite[Thm.~7.6]{Rub26}. Hence for every fixed $\beta_0$ the single-copy rate on $\mathcal C(\beta_0)$ is $\Theta(m^3/\eps^2)$ on the window of (i).
\item[(iii)] For $\beta_0\le0.1$ and $\eps\ge\tfrac12(1+5\beta_0)\beta_0$, the fixed output $I/2^m$ is $\eps$-accurate on all of $\mathcal C(\beta_0)$, so no copies are needed.
\end{enumerate}
\end{theorem}

\begin{proof}
(i) The hard family of Theorem~\ref{thm:kcopy-mixed} is $\{\rho_\theta:\|\theta\|_2\le120\sqrt3\,\eps\}$, which is contained in $\mathcal C(\beta_0)$ exactly when $120\sqrt3\,\eps\le\beta_0$; a lower bound on a subclass is a lower bound on the class. (ii) $\|\Gamma(\rho_\theta)\|_\op=\max_j\tanh\lambda_j\le\tanh\|\theta\|_2\le\tanh\beta_0$ by \cite[Eq.~(2.3)]{Rub26} and $\lambda_j\le\|\lambda\|_2=\|\theta\|_2$. (iii) By the remark following \cite[Prop.~5.4]{Rub26}, $T(\rho_\theta,\rho_0)\le\tfrac12(1+5r)\|\theta\|_2$ for $\|\theta\|_2\le r\le0.1$; take $r=\beta_0$.
\end{proof}

\paragraph{Discussion.} In \cite{CMMP26} a family of bosonic Gaussian states is called classical to learn when it can be learned with non-entangled measurements at the sample complexity $\Theta(m^2/\eps^2)$ of its classical analogue, the $2m$-variate Gaussian distribution; warm states, with a constant number of thermal photons per mode, are classical to learn, and cold states are not. The criterion presupposes a classical analogue, and fermionic Gaussian states have none: their phase-space representation is Grassmann-valued rather than a probability distribution, and the covariance is bounded by $\|\Gamma\|_\op\le1$, so no parameter plays the role of the thermal-photon number $\nu\to\infty$. Theorem~\ref{prop:nowarm} makes the consequence quantitative. The fermionic analogue of warming a state is lowering $\beta_0$, and as $\beta_0$ decreases the single-copy rate on $\mathcal C(\beta_0)$ does not decrease: it is $m^3$ for every $\beta_0$, with a lower-bound constant that does not depend on $\beta_0$, while the upper-bound constant of (ii) grows like $(1-\tanh\beta_0)^{-1}$. What changes with $\beta_0$ is only the window of accuracies on which the question is nontrivial, which closes linearly, $\eps\lesssim\beta_0$, because the whole class collapses to $I/2^m$. Between the window of (i) and the trivial regime of (iii) lies a constant-factor gap in $\eps/\beta_0$ in which the rate is not determined. In the bosonic case the analogous window never closes: warm states remain $\eps$-distinguishable from one another at every scale, and their rate drops to $m^2$. The contrast is the answer, for fermionic Gaussian states, to the question of \cite[Sec.~1.4]{CMMP26}: the crossover does not generalise; there is a cold regime only.

\paragraph{One-sided covariance bounds.} Both \cite{CMMP26} and \cite{Rub26} convert covariance error into trace-distance error with constants depending only on the distance from the pure boundary, and in both a hypothesis at one endpoint suffices. For bosons \cite[Thm.~7]{CMMP26}: if the symplectic eigenvalues of $\bar\Sigma$ are at least $\nu_->\tfrac12$ and $\Sigma'\succeq\bar\Sigma$, then $T(\rho(\Sigma'),\rho(\bar\Sigma))\le\|\bar\Sigma^{-1/2}(\Sigma'-\bar\Sigma)\bar\Sigma^{-1/2}\|_F/(2\sqrt{2\gamma_-})$ with $\gamma_-=1-1/(2\nu_-)$. For fermions \cite[Cor.~7.2]{Rub26}: if $\|\Gamma\|_\op\le1-c$ and $\Gamma'$ is any covariance matrix, then $\|\rho_\Gamma-\rho_{\Gamma'}\|_1\le\sqrt{2/c}\,\|\Gamma-\Gamma'\|_F$, improving to $(2c(2-c))^{-1/2}$ when $\|\Gamma'\|_\op\le1-c$ as well. For fermions one mixed endpoint gives the larger constant $\sqrt{2/c}$ and two mixed endpoints the smaller one; the bosonic hypothesis is an order relation between the two covariances, a different notion of one-sidedness, and the two should not be identified. In both statistics no dimension-free constant exists on the full class \cite[Prop.~7.4]{Rub26}, which is the source of the $m^4$ in Table~\ref{tab:rates}.

\section{Near the vacuum: the exterior-power embedding}\label{sec:vacuum}

We now turn to the near-vacuum class, where the lower bound comes from the embedding of qudit tomography of \cite[Sec.~9]{Rub26} and \cite[Sec.~4]{CMMP26}, run through Lemma~\ref{lem:fd}; this section recomputes its constants for exterior powers.

\subsection{The encoding and its constants}

Let $\Fock=\bigoplus_{n=0}^m\Fock_n$ be the Fock space with $\Fock_n$ the $n$-particle sector, spanned by $a^\dagger_{i_1}\cdots a^\dagger_{i_n}|0\rangle$, $i_1<\dots<i_n$. Let $\Lam^n(\C^m)\subset(\C^m)^{\otimes n}$ be the exterior power, $A_n$ the antisymmetriser projecting onto it, $\Lam^0=\C$ and $\Lam^n=0$ for $n>m$. The map $V_n\colon e_{i_1}\wedge\cdots\wedge e_{i_n}\mapsto a^\dagger_{i_1}\cdots a^\dagger_{i_n}|0\rangle$ is a unitary $\Lam^n(\C^m)\to\Fock_n$, and we write $\Fock_n=\Lam^n(\C^m)$ when no confusion arises. All states in this section are even, so $b$-fold tensor products of Fock spaces are unambiguous. For an operator $X$ on $\C^m$ put $\Lam^n(X)=A_nX^{\otimes n}A_n=X^{\otimes n}A_n$ and
\[
\Gamma_{\rm a}(X)=\bigoplus_{n=0}^m\Lam^n(X),\qquad\Tr\Lam^n(X)=e_n(x_1,\dots,x_m),\qquad\Tr\Gamma_{\rm a}(X)=\prod_i(1+x_i)=\det(I+X),
\]
with $x_i$ the eigenvalues of $X$ and $e_n$ the elementary symmetric polynomial; under $V_n$, $\Lam^n(X)$ is the $n$-particle block of the second quantisation of $X$. For a density matrix $w$ on $\C^m$ with eigenvalues $\lambda_1,\dots,\lambda_m$ define
\begin{equation}\label{eq:encoding}
D(w)=\det(I+w/2),\qquad\rho^{\rm F}_w=\frac{\Gamma_{\rm a}(w/2)}{D(w)} .
\end{equation}
In the language of \cite{CMMP26}, $w/2$ is the fugacity matrix of $\rho^{\rm F}_w$ (they use $\sigma/4$ for bosons; the factor changes only constants).

\begin{lemma}[the class]\label{lem:class}
\begin{enumerate}
\item[(i)] $\tfrac32\le D(w)\le e^{1/2}$.
\item[(ii)] $\rho^{\rm F}_w$ is a number-conserving fermionic Gaussian state. In the eigenbasis of $w$ it is the product over modes of $(1-\nu_i)|0\rangle\langle0|+\nu_i|1\rangle\langle1|$ with occupations $\nu_i=\lambda_i/(2+\lambda_i)\in[0,\tfrac13]$; its occupation matrix is $\mathsf N(w)=(w/2)(I+w/2)^{-1}$, with $\tfrac13\le\Tr\mathsf N(w)<\tfrac12$. If $w\preceq2I/m$ then every $\nu_i\le1/(m+1)$, and the covariance normal-form values $1-2\nu_i$ are at least $1-2/(m+1)$.
\item[(iii)] $\Pi_1\rho^{\rm F}_w\Pi_1=w/(2D(w))$ on $\Fock_1=\C^m$, and $a(w):=\Tr(\Pi_1\rho^{\rm F}_w)=1/(2D(w))\in[\tfrac12e^{-1/2},\tfrac13]$.
\item[(iv)] $\|\rho^{\rm F}_w-\rho^{\rm F}_{w'}\|_1\le\tfrac83\|w-w'\|_1$.
\end{enumerate}
\end{lemma}

\begin{proof}
(i) $D(w)=\prod_i(1+\lambda_i/2)$. Expanding the product, every term is nonnegative and the terms of degree at most one sum to $1+\tfrac12\sum_i\lambda_i=\tfrac32$; and $1+x\le e^x$ gives $D\le e^{\sum_i\lambda_i/2}=e^{1/2}$.

(ii) In the eigenbasis of $w$, $\Gamma_{\rm a}(\operatorname{diag}(x))$ acts on the occupation basis $|n_1,\dots,n_m\rangle$ as $\prod_ix_i^{n_i}$, so $\Gamma_{\rm a}(w/2)=\bigotimes_i(|0\rangle\langle0|+(\lambda_i/2)|1\rangle\langle1|)$ up to the passive rotation diagonalising $w$, and dividing by $D(w)$ gives the product of single-mode states with $\nu_i=(\lambda_i/2)/(1+\lambda_i/2)=\lambda_i/(2+\lambda_i)$. A product of single-mode diagonal states is $\prod_i\tfrac12(I+\mu_iZ_i)$ with $\mu_i=1-2\nu_i$, which is Gaussian \cite[Sec.~2]{Rub26}, and a passive rotation preserves Gaussianity and particle number. Since $0\le\lambda_i\le1$, $\nu_i\le\tfrac13$. The function $f(\lambda)=\lambda/(2+\lambda)$ is concave on $[0,1]$, so on the simplex $\sum_if(\lambda_i)$ is at most $mf(1/m)=1/(2+1/m)<\tfrac12$ by Jensen and at least its value at a vertex, $f(1)=\tfrac13$. If $w\preceq2I/m$ then $\lambda_i\le2/m$ and $\nu_i\le(2/m)/(2+2/m)=1/(m+1)$.

(iii) $\Lam^1(w/2)=w/2$, so $a(w)=\Tr(w/2)/D(w)=1/(2D(w))$, and (i) gives the range.

(iv) By symmetry assume $D(w)\ge D(w')$. Telescoping gives
\[
\|X^{\otimes n}-Y^{\otimes n}\|_1\le n\|X-Y\|_1\max(\|X\|_1,\|Y\|_1)^{n-1},
\]
and compression by $A_n$ is trace-norm contractive, so $\|\Lam^n(w/2)-\Lam^n(w'/2)\|_1\le n2^{-n}\|w-w'\|_1$. Summing over $n\ge1$, $\sum_nn2^{-n}=2$, gives $\|\Gamma_{\rm a}(w/2)-\Gamma_{\rm a}(w'/2)\|_1\le2\|w-w'\|_1$. For positive trace-class $A,B$ with traces $\alpha\ge\beta>0$, $\|A/\alpha-B/\beta\|_1\le(\|A-B\|_1+|\alpha-\beta|)/\alpha\le2\|A-B\|_1/\alpha$; with $\alpha=D(w)\ge\tfrac32$ this gives $\tfrac43\cdot2\|w-w'\|_1$.
\end{proof}

\begin{lemma}[readout]\label{lem:readout}
Let $0<\eps\le1/100$ and let $\hat\rho$ be any density matrix on $\Fock$ with $T(\hat\rho,\rho^{\rm F}_w)\le\eps$. Put $B=\Pi_1\hat\rho\Pi_1$ and $b=\Tr B$. Then $b>0$, $\hat w=B/b$ is a density matrix on $\C^m$, and
\[
T(\hat w,w)\le4D(w)\,\eps\le4e^{1/2}\eps<6.6\,\eps .
\]
\end{lemma}

\begin{proof}
Let $A=\Pi_1\rho^{\rm F}_w\Pi_1=w/(2D)$ and $a=\Tr A=1/(2D)\ge\tfrac12e^{-1/2}>0.30$. Compression by $\Pi_1$ is trace-norm contractive, so $\|A-B\|_1\le2\eps$ and $|a-b|\le2\eps$; hence $b\ge0.30-0.02>0$, and $B\succeq0$ makes $\hat w$ a density matrix. Then
\[
\|\hat w-w\|_1\le\Big\|\frac Bb-\frac Ba\Big\|_1+\frac{\|B-A\|_1}{a}=\frac{|a-b|}{a}+\frac{\|A-B\|_1}{a}\le\frac{4\eps}{a}=8D(w)\eps ,
\]
using $\|B/b-B/a\|_1=b|1/b-1/a|=|a-b|/a$. Halving gives the claim, and $4e^{1/2}=6.5949\ldots<6.6$. The hypothesis $\eps\le1/100$ is used only to ensure $b>0$. (The bosonic Lemma~4 of \cite{CMMP26} obtains $16\eps$ because it divides by the crude bound $\hat p>1/8$ instead of by $p_\sigma$; the difference is bookkeeping.)
\end{proof}

\subsection{The sampler and one batch}

\paragraph{One trial.} Sample $L\in\{0,1,2,\dots\}$ with $\Prob(L=\ell)=2^{-\ell-1}$, so $\E L=1$ and $\operatorname{Var}L=2$. If $L=0$, the trial succeeds and outputs the vacuum, consuming no copy. If $L\ge1$, take $L$ fresh copies of $w$ and measure $\{A_L,I-A_L\}$ on $(\C^m)^{\otimes L}$; on outcome $A_L$ the trial succeeds and outputs the post-measurement state, an element of $\Lam^L(\C^m)=\Fock_L$; on $I-A_L$ it fails and the copies are discarded. For $L>m$, $A_L=0$ and the trial always fails; this needs no special treatment.

\begin{lemma}[one trial]\label{lem:trial}
For fixed $L=\ell$ the unnormalised post-measurement state is $\Lam^\ell(w)$, with success probability $e_\ell(\lambda)$. Averaged over $L$, the subnormalised successful output is
\[
\sum_{\ell\ge0}2^{-\ell-1}\Lam^\ell(w)=\tfrac12\Gamma_{\rm a}(w/2)=q_w\,\rho^{\rm F}_w,\qquad q_w:=\tfrac12D(w)\in[\tfrac34,\tfrac12e^{1/2}],
\]
so a trial succeeds with probability $q_w\ge\tfrac34$ and, conditioned on success and after discarding $L$, outputs exactly $\rho^{\rm F}_w$.
\end{lemma}

\begin{proof}
$w^{\otimes\ell}$ commutes with $A_\ell$, so $A_\ell w^{\otimes\ell}A_\ell=\Lam^\ell(w)$ with trace $e_\ell(\lambda)$, and $\Lam^\ell(cX)=c^\ell\Lam^\ell(X)$. Summing, $\sum_\ell2^{-\ell-1}\Lam^\ell(w)=\tfrac12\sum_\ell\Lam^\ell(w/2)=\tfrac12\Gamma_{\rm a}(w/2)$, whose trace is $\tfrac12D(w)$; Lemma~\ref{lem:class}(i) gives the range.
\end{proof}

\paragraph{One batch.} Fix a request for $b$ copies, $1\le b\le k$. Sample $2b$ independent lengths $L_1,\dots,L_{2b}$, set $X=\sum_aL_a$ and $t=\max\{1,X\}$. Take $t$ fresh copies of $w$ and run the $2b$ trials on disjoint groups, discarding the extra copy when all lengths vanish. If at least $b$ trials succeed, apply the learner's requested POVM $M^{(h)}=\{M^{(h)}_y\}$ to the first $b$ successful outputs and pass only the outcome $y$ to the learner; otherwise discard the batch and repeat with fresh copies, without advancing the learner. Every quantum system is discarded at the end of each batch.

\begin{lemma}[one batch]\label{lem:batch}
\begin{enumerate}
\item[(i)] For each preselected length vector the batch is one POVM on $t$ fresh copies of $w$, with outcomes indexed by a success pattern $P\subset\{1,\dots,2b\}$ and, when $|P|\ge b$, an outcome $y$. Writing $S\subset P$ for the first $b$ elements of $P$, $\bar A_a=A_{L_a}$ on the group of trial $a$ (the scalar $1$ when $L_a=0$), and $V=\bigotimes_{a\in S}V_{L_a}$,
\begin{align*}
E_{P,y}&=\Big[\textstyle\bigotimes_{a\in S}\bar A_a\Big]V^\dagger\tilde M^{(h)}_yV\Big[\textstyle\bigotimes_{a\in S}\bar A_a\Big]\otimes\bigotimes_{a\in P\setminus S}\bar A_a\otimes\bigotimes_{a\notin P}(I-\bar A_a)\qquad(|P|\ge b),\\
E_P&=\bigotimes_{a\in P}\bar A_a\otimes\bigotimes_{a\notin P}(I-\bar A_a)\qquad(|P|<b),
\end{align*}
where $\tilde M^{(h)}_y=\Tr_{\rm anc}[(I\otimes\eta^{1/2})M^{(h)}_y(I\otimes\eta^{1/2})]$ absorbs the learner's fresh ancillas in state $\eta$; when all $L_a=0$ the block is a single copy with the trivial POVM $E_y=\Tr[\tilde M^{(h)}_y|0\rangle\langle0|^{\otimes b}]\,I$. The block size $t$ and the POVM are fixed before the $t$ copies are touched. (The retry logic uses previous batches' success flags, which Lemma~\ref{lem:fd} permits; only the learner is blinded to them.)
\item[(ii)] The batch succeeds with probability $p_b(w)\ge\tfrac12$.
\item[(iii)] Conditioned on success, the $b$ states handed to the POVM are exactly $(\rho^{\rm F}_w)^{\otimes b}$; hence the classical record seen by the learner has exactly the law of a run on fresh copies of $\rho^{\rm F}_w$.
\item[(iv)] $\E t\le1+2b\le3b$ and $\E t^2\le1+4b+4b^2\le9b^2$, so with $w_m(t)=t\sqrt{\min\{t,m^2\}}=\min\{t^{3/2},mt\}$,
\[
\E\,w_m(t)\le\min\{(\E t^2)^{3/4},\,m\E t\}\le\min\{9^{3/4}b^{3/2},\,3mb\}\le5.2\,b\sqrt{\min\{b,m^2\}} .
\]
\end{enumerate}
\end{lemma}

\begin{proof}
(i) Each element is positive; summing $E_{P,y}$ over $y$ gives $\prod_{a\in P}\bar A_a\prod_{a\notin P}(I-\bar A_a)$ since $\sum_y\tilde M^{(h)}_y=I$, and $\sum_P\prod_{a\in P}\bar A_a\prod_{a\notin P}(I-\bar A_a)=I$. The lengths are sampled, and $M^{(h)}$ chosen from the learner's classical history $h$, before any copy is measured.

(ii) Let $F$ be the number of failed trials among the $2b$. Trials are independent with failure probability $1-q_w\le\tfrac14$ by Lemma~\ref{lem:trial}, so $\E F\le b/2$; the batch fails iff $F\ge b+1$, and Markov gives $1-p_b(w)\le\E F/(b+1)\le b/(2(b+1))\le\tfrac12$.

(iii) As in \cite[proof of Thm.~4]{CMMP26}: for a success pattern with $s\ge b$ successes the subnormalised state on the successful outputs, after averaging over the lengths, is $q_w^s(1-q_w)^{2b-s}(\rho^{\rm F}_w)^{\otimes s}$ by Lemma~\ref{lem:trial} and independence; keeping the first $b$ and summing over patterns gives $p_b(w)(\rho^{\rm F}_w)^{\otimes b}$. The identity holds after averaging over the lengths, not conditionally on $t$, so the lengths and success flags are not passed to the learner, whose choices depend only on its own outcomes and randomness; repeated batches use fresh copies.

(iv) $\E X=2b$, $\operatorname{Var}X=4b$, $\E X^2=4b+4b^2$; $t\le1+X$ and $t^2\le1+X^2$ in both cases $X=0$ and $X\ge1$; and $9b^2-(4b^2+4b+1)=(5b+1)(b-1)\ge0$. Jensen for the concave $x\mapsto x^{3/4}$ gives $\E t^{3/2}\le(\E t^2)^{3/4}\le9^{3/4}b^{3/2}<5.2\,b^{3/2}$, and $m\E t\le3mb$.
\end{proof}

\subsection{The theorem}

\begin{theorem}[fermions near the vacuum, $k$ copies]\label{thm:vacuum}
Let $\eps_{\rm fd},d_0$ be the constants of Lemma~\ref{lem:fd}, $m_0=\max\{d_0,2\}$ and $\eps_0=\min\{1/100,\eps_{\rm fd}/6.6\}$. Let $m\ge m_0$, $0<\eps<\eps_0$ and $k\ge1$. Any $k$-copy protocol that uses at most $N$ copies and, for every density matrix $w\preceq2I/m$ on $\C^m$, outputs $\hat\rho$ on $\Fock$ with $T(\hat\rho,\rho^{\rm F}_w)\le\eps$ with probability at least $2/3$, satisfies
\[
N=\Omega\Big(\frac{m^3}{\eps^2\min\{\sqrt k,m\}}\Big).
\]
The same holds for protocols required to work only on $\{w:\|w-I/m\|_\op\le L\eps/m\}$ with $L=6.6\,C_{\rm fd}$, by Lemma~\ref{lem:fd}$'$. The class $\{\rho^{\rm F}_w:w\preceq2I/m\}$ is learnable with $O(m^3/\eps^2)$ single-copy and $\Theta(m^2/\eps^2)$ collective copies.
\end{theorem}

\begin{proof}
We build from the protocol $\mathcal A$ a finite-dimensional protocol for $w\preceq2I/m$ satisfying the hypotheses of Lemma~\ref{lem:fd} with $d=m$.

\emph{Exact simulation.} Whenever $\mathcal A$ requests a block of $b_i$ Gaussian copies, run batches of Lemma~\ref{lem:batch} until one succeeds and pass the POVM outcome to $\mathcal A$. By Lemma~\ref{lem:batch}(iii) the transcript of $\mathcal A$ inside the simulation has exactly the law of $\mathcal A$ on fresh copies of $\rho^{\rm F}_w$, so the final estimate $\hat\rho$ is within $\eps$ of $\rho^{\rm F}_w$ with probability at least $2/3$.

\emph{Cost of one request.} Let $R_b$ be the index of the first successful batch for a request of size $b$ and $t_\ell$ the sizes of the batches in the infinite independent sequence of which only the first $R_b$ are performed. Since $\{R_b\ge\ell\}$ depends only on batches $1,\dots,\ell-1$ and is independent of $t_\ell$,
\[
\E_w\Big[\sum_{\ell=1}^{R_b}t_\ell\Big]=\frac{\E t}{p_b(w)}\le2\E t\le6b,\qquad
\E_w\Big[\sum_{\ell=1}^{R_b}w_m(t_\ell)\Big]=\frac{\E w_m(t)}{p_b(w)}\le10.4\,b\sqrt{\min\{b,m^2\}} ,
\]
by Lemma~\ref{lem:batch}(ii),(iv).

\emph{Total cost.} Let $H_i$ be the classical information available after $\mathcal A$ has chosen its $i$-th block, including the simulation's own earlier randomness, so that $b_i$ and the POVM are fixed given $H_i$; let $S_i$ be the number of copies of $w$ used for request $i$ including failed batches and $W_i$ the summed weighted cost; set $b_i=S_i=W_i=0$ after $\mathcal A$ stops, and sum $i$ from $1$ to $N$, there being at most $N$ requests since every $b_i\ge1$ and $\sum_ib_i\le N$ on every run. The simulation of request $i$ uses fresh copies and fresh randomness, so the single-request bounds hold conditionally on $H_i$, and the law of total expectation gives, uniformly over $w\preceq2I/m$,
\begin{align*}
\E_wS&\le6\,\E_w\sum_ib_i\le6N,\\
\E_w\sum_jw_m(t_j)&=\E_w\sum_iW_i\le10.4\,\E_w\sum_ib_i\sqrt{\min\{b_i,m^2\}}\le10.4\,N\sqrt{\min\{k,m^2\}} ,
\end{align*}
where $S=\sum_iS_i$, $j$ indexes all batches performed, failed ones included, and the last step uses $b_i\le k$.

\emph{Deterministic budget.} Put $M_0=72N$. Before each batch its size $t$ is already known; if performing it would push the total number of copies used above $M_0$, abort without taking those copies. Coupling the capped and exact simulations on the same randomness, an abort implies $S>M_0$ for the exact simulation, so $\Prob_w(\mathrm{abort})\le\E_wS/M_0\le1/12$. The capped run performs a prefix of the exact run's batches and all weights are nonnegative, so its expected weighted cost is at most $10.4\,N\sqrt{\min\{k,m^2\}}$. When no abort occurs the two runs coincide, so the capped protocol produces $\hat\rho$ within $\eps$ of $\rho^{\rm F}_w$ with probability at least $1-\tfrac13-\tfrac1{12}=\tfrac7{12}>\tfrac12$.

\emph{Readout.} Return $\hat w=\Pi_1\hat\rho\Pi_1/\Tr(\Pi_1\hat\rho)$, or a fixed state on abort or if the denominator vanishes. By Lemma~\ref{lem:readout}, whenever $\hat\rho$ is within $\eps\le1/100$ of $\rho^{\rm F}_w$, $\hat w$ is within $6.6\eps$ of $w$. So the capped protocol learns every $w\preceq2I/m$ to trace distance $6.6\eps$ with probability at least $7/12$.

\emph{Black box.} The capped protocol has deterministic budget $M_0$, chooses each block size and POVM before touching the copies (Lemma~\ref{lem:batch}(i)), keeps no quantum memory between blocks, and satisfies $\sum_jt_j\le M_0$ almost surely. With $m\ge d_0$ and $6.6\eps<\eps_{\rm fd}$, Lemma~\ref{lem:fd} gives
\[
\Omega\Big(\frac{m^3}{(6.6\eps)^2}\Big)\le\sup_{w\preceq2I/m}\E_{w,\rm capped}\sum_jw_m(t_j)\le10.4\,N\sqrt{\min\{k,m^2\}} ,
\]
and $\sqrt{\min\{k,m^2\}}=\min\{\sqrt k,m\}$ gives the bound. For the local version, every state in play is some $\rho^{\rm F}_w$ with $w$ in the promise set, because the sampler is exact, and the readout is at accuracy $6.6\eps$, so Lemma~\ref{lem:fd}$'$ applies with radius $C_{\rm fd}\cdot6.6\eps/m$.

\emph{Upper bounds.} Single-copy: measure the particle number on each copy; outcome $1$ has probability $a(w)>0.30$ and leaves exactly $w$ on $\Fock_1=\C^m$; the covariant single-copy qudit learner of \cite[proof of Cor.~9.5]{Rub26} learns $w$ to trace distance $\eps'$ with $O(m^3/\eps'^2)$ copies; output $\rho^{\rm F}_{\hat w}$, within $\tfrac83\eps'$ of $\rho^{\rm F}_w$ by Lemma~\ref{lem:class}(iv), and take $\eps'=3\eps/8$. Collective: $O(m^2/\eps^2)$ by \cite[Thm.~4.2]{CFGLMWW26} for all fermionic Gaussian states, and $\Omega(m^2/\eps^2)$ on the class by the embedding with quantum memory, exactly as in \cite[proof of Cor.~9.5]{Rub26}, with the constants of Lemma~\ref{lem:readout}.
\end{proof}

\begin{remark}[what changed from the bosonic proof]
Compared with \cite[proof of Thm.~4]{CMMP26}: the fugacity is $w/2$ instead of $\sigma/4$, so $\E L=1$ and $\operatorname{Var}L=2$ instead of $1/3$ and $4/9$, and the constants $5b/3$, $7b^2/3$, $10N/3$, $40N$ become $3b$, $9b^2$, $6N$, $72N$; the symmetric projector is replaced by the antisymmetriser, the generating function $\sum_\ell x^\ell\Tr\Lam^\ell(w)=\det(I+xw)$ replaces $\det(I-x\sigma)^{-1}$, and $e_\ell\le1$ replaces $h_\ell\le1$; the trial success probability $q_w=D(w)/2\ge3/4$ coincidentally has the same lower bound as their $3/(4\det(I-\sigma/4))$, so the Markov step is unchanged; the readout constant is $6.6$ instead of $16$. Nothing else in the argument refers to the symmetric structure.
\end{remark}

\subsection{A transfer lemma}

The proof of Theorem~\ref{thm:vacuum} uses Lemma~\ref{lem:fd} only at its last line. Everything before it is a statement about protocols, which we record so that other qudit lower bounds in the block model can be moved to the Gaussian class without repeating the argument.

\begin{proposition}[transfer]\label{prop:transfer}
Let $\mathcal A$ be a $k$-copy protocol on $\Fock$ with at most $N$ copies, and let $\mathcal C\subset\mathcal D(\C^m)$. There is a protocol $\mathcal B$ on $\C^m$ in the model of Lemma~\ref{lem:fd}, with deterministic budget $72N$ and
\[
\sup_{w\in\mathcal C}\E_w\sum_jw_m(t_j)\le10.4\,N\sqrt{\min\{k,m^2\}},
\]
such that for every $w\in\mathcal C$ there is a coupling under which, with probability at least $11/12$, $\mathcal B$ outputs the classical output of a run of $\mathcal A$ on fresh copies of $\rho^{\rm F}_w$, and otherwise outputs a fixed symbol. The map $w\mapsto\rho^{\rm F}_w$ satisfies $\tfrac1{6.6}T(w,w')\le T(\rho^{\rm F}_w,\rho^{\rm F}_{w'})\le\tfrac83T(w,w')$, and the same construction with symmetric powers and the constants of \cite[Sec.~9]{Rub26} or \cite[Sec.~4]{CMMP26} gives the bosonic statement.
\end{proposition}

\begin{proof}
$\mathcal B$ is the capped simulation of the proof of Theorem~\ref{thm:vacuum} without the readout step; the budget, weighted cost and abort probability are as computed there, and the coupling is the one used there. The upper Lipschitz bound is Lemma~\ref{lem:class}(iv). For the lower one, run the computation of Lemma~\ref{lem:readout} with $\hat\rho=\rho^{\rm F}_{w'}$: the denominator is $b=1/(2D(w'))>0$ automatically, so the hypothesis $\eps\le1/100$ is not needed and the readout, which is exactly $w'$, satisfies $T(w,w')\le4D(w)\,T(\rho^{\rm F}_w,\rho^{\rm F}_{w'})\le6.6\,T(\rho^{\rm F}_w,\rho^{\rm F}_{w'})$.
\end{proof}

Thus any lower bound of the form ``every protocol in the model of Lemma~\ref{lem:fd} that accomplishes a task on $\mathcal C$ with probability at least $1/2$ has $\sup_{w\in\mathcal C}\E_w\sum_jw_m(t_j)\ge\Lambda$'' yields $N\ge\Lambda/(10.4\min\{\sqrt k,m\})$ for $k$-copy protocols accomplishing the transported task on $\{\rho^{\rm F}_w:w\in\mathcal C\}$ with probability $2/3$, provided the task is stable under the bi-Lipschitz map with the stated constants. Tomography is the case $\Lambda=\Omega(m^3/\eps^2)$. For the low-rank tomography bounds of Nayak and Zhou \cite{NZ26}, and for single-copy testing bounds, the hypothesis is that the qudit bound be available in the weighted block form; we return to this in Section~\ref{sec:open}.

\subsection{The \texorpdfstring{$k$}{k}-copy rate of the cold number-conserving classes, for every \texorpdfstring{$k$}{k}}\label{sec:kcopy-rate}

The embedding also carries upper bounds, in the other direction: a $k$-copy protocol for qudits becomes a $k$-copy protocol for the Gaussian class by reading the qudit off the one-particle sector. Two things make this determine the $k$-copy rate for every $k$, and on a class much larger than the encoded one. First, the qudit side is now settled for every block size: measurements on at most $t$ copies at a time learn every $d$-dimensional state to trace-norm error $\eps$ from $O(\max\{d^3/(\sqrt t\,\eps^2),d^2/\eps^2\})$ copies, by independent applications of the debiased Keyl measurement of Pelecanos, Spilecki and Wright \cite{PSW26} and by the reduction of Pelecanos, Spilecki, Tang and Wright \cite{PSTW26}; and Nayak and Zhou \cite[Thm.~1.2]{NZ26} prove the rank-dependent form $O((dr/\eps^2)\max\{1,r/\sqrt t\})$ for states of rank at most $r$, for every $t\ge1$ and every $0<\eps\le1$, by a nonadaptive protocol that measures blocks of exactly $\min\{t,r^2\}$ copies. It is the last form, at $r=d$, that we use. The range restriction $t\le c_1(\sqrt d/\eps)^{1/5}$ of the earlier protocol of Chen, Li and Liu \cite[Thm.~5.7]{CLL24} is gone. Second, the one-particle readout does not need a fixed weight in the one-particle sector. If that weight is small the state is close to the vacuum, and the accuracy to which the one-particle state must be learned relaxes in exact proportion; what makes this quantitative is a trace-distance bound that is linear in the trace norm of the fugacity difference and uniform in the number of modes and in the fugacity trace (Lemma~\ref{lem:coldclass}(ii)).

For $C>0$ and $0<\delta\le1$ let
\begin{align*}
\mathcal N^{\rm F}(C)&=\Big\{\rho_X:=\frac{\Gamma_{\rm a}(X)}{\det(I+X)}:\ X\succeq0\text{ on }\C^m,\ \tr X\le C\Big\},\\
\mathcal N^{\rm B}(C,\delta)&=\Big\{\rho_X:=\det(I-X)\,\Gamma_{\rm s}(X):\ X\succeq0\text{ on }\C^m,\ \|X\|_\op\le1-\delta,\ \Tr X(I-X)^{-1}\le C\Big\},
\end{align*}
where $\Gamma_{\rm s}(X)=\bigoplus_{\ell\ge0}\operatorname{Sym}^\ell(X)$ is the bosonic second quantisation, $\operatorname{Sym}^\ell(X)=S_\ell X^{\otimes\ell}S_\ell$ with $S_\ell$ the symmetriser. $\mathcal N^{\rm F}(C)$ is the set of number-conserving fermionic Gaussian states, the Gibbs states $\propto\exp(-\sum_{jk}h_{jk}a_j^\dagger a_k)$ and their limits, whose fugacity matrix $X=e^{-h}$ has trace at most $C$; its occupation matrix is $X(I+X)^{-1}$, so the expected particle number is at most $C$, and $\{\rho^{\rm F}_w\}=\{\rho_{w/2}\}\subset\mathcal N^{\rm F}(C)$ as soon as $C\ge\tfrac12$. $\mathcal N^{\rm B}(C,\delta)$ is the set of passive bosonic Gaussian states \cite[Sec.~4]{CMMP26} with mean photon number $\Tr X(I-X)^{-1}$ at most $C$ and fugacity eigenvalues at most $1-\delta$. It contains the encoded class $\{\rho_\sigma\}$ of \cite{CMMP26}, fugacity $\sigma/4\preceq I/(2m)$, for $C\ge\tfrac13$, $\delta\le\tfrac34$ and $m\ge2$; and it contains, for every constant $c>0$ and $m\ge2c$, every passive state with covariance $\Sigma\preceq(\tfrac12+c/m)I$, with $C=c$ and $\delta=\tfrac12$, since such a state has at most $c/m$ photons in each normal mode, hence mean photon number at most $c$ and fugacity eigenvalues $x_j=n_j/(1+n_j)\le c/m\le\tfrac12$. This is exactly the passive part of the cold promise class of \cite[Thm.~2]{CMMP26}, for any value of their implied constant. In both statistics we write $\Gamma(X)$ for the second quantisation, $Z(X)=\Tr\Gamma(X)$, so that $\rho_X=\Gamma(X)/Z(X)$; on both classes $\tr X\le C$ and
\[
Z(X)\le Z_0:=e^{C},
\]
since $\det(I+X)\le e^{\tr X}$ and $\det(I-X)^{-1}=e^{-\sum_j\log(1-x_j)}\le e^{\sum_jx_j/(1-x_j)}$; and we put $L:=2$ for fermions and $L:=4/\delta$ for bosons. (The photon-number bound already forces $x_j/(1-x_j)\le C$, hence $\|X\|_\op\le C/(1+C)$, so $\delta$ may always be taken to be $1/(1+C)$; we keep it as a separate parameter to track the constants.)

\begin{lemma}[the cold classes]\label{lem:coldclass}
Let $X\succeq0$ on $\C^m$, $s=\tr X$, and $\tau=X/s$ when $s>0$; for bosons assume $\|X\|_\op<1$.
\begin{enumerate}
\item[(i)] The particle-number law of $\rho_X$ has $P_0=1/Z(X)$ and $P_1=s/Z(X)$, so $s=P_1/P_0$; and on outcome $1$ the post-measurement state is exactly $\tau$ on $\Fock_1=\C^m$. On $\mathcal N^{\rm F}(C)$ and $\mathcal N^{\rm B}(C,\delta)$, $P_0\ge e^{-C}$ and $P_1\ge se^{-C}$.
\item[(ii)] Let $X'\succeq0$, $Y=X'-X$ and $X_u=X+uY$. Then
\[
\|\rho_X-\rho_{X'}\|_1\le2\int_0^1\Tr\big((I\pm X_u)^{-1}|Y|\big)\,du\le2\,\|X-X'\|_1\sup_{0\le u\le1}\|(I\pm X_u)^{-1}\|_\op,
\]
with the upper sign for fermions and the lower sign for bosons, where for bosons $\|X'\|_\op<1$ is assumed. Hence $\|\rho_X-\rho_{X'}\|_1\le2\|X-X'\|_1$ for all fermionic $X,X'\succeq0$, and $\|\rho_X-\rho_{X'}\|_1\le(2/\delta')\|X-X'\|_1$ for bosonic $X,X'$ with $\|X\|_\op,\|X'\|_\op\le1-\delta'$. In particular $\|\rho_X-|0\rangle\langle0|\|_1\le L\,s$ on both classes.
\end{enumerate}
\end{lemma}

\begin{proof}
(i) The $\ell$-particle block of $\Gamma(X)$ is $\Lam^\ell(X)$ or $\operatorname{Sym}^\ell(X)$, with traces $e_\ell(x)$ or $h_\ell(x)$; $\ell=0$ gives $1$ and $\ell=1$ gives $X$ itself, of trace $s$. The bound $Z\le e^C$ was shown above.

(ii) Write $\Gamma_\ell(X)=\Pi_\ell X^{\otimes\ell}\Pi_\ell$ for the $\ell$-particle block, $\Pi_\ell=A_\ell$ or $S_\ell$, and $\rho_\ell(u)=\Gamma_\ell(X_u)/Z(X_u)$, a smooth function of $u$ on a neighbourhood of $[0,1]$ with values in the finite-dimensional block, since $Z$ is smooth and positive there (for bosons $\|X_u\|_\op\le\max\{\|X\|_\op,\|X'\|_\op\}<1$ by convexity). Since the blocks are orthogonal, $\|\rho_X-\rho_{X'}\|_1=\sum_\ell\|\rho_\ell(1)-\rho_\ell(0)\|_1\le\sum_\ell\int_0^1\|\rho_\ell'(u)\|_1\,du=\int_0^1\sum_\ell\|\rho_\ell'(u)\|_1\,du$, the exchange by Tonelli. Now
\[
\rho_\ell'(u)=\frac{\Gamma_\ell'(u)}{Z(X_u)}-\rho_\ell(u)\,\frac{d}{du}\log Z(X_u),\qquad
\Gamma_\ell'(u)=\Pi_\ell\Big[\sum_{i=1}^\ell X_u^{\otimes(i-1)}\otimes Y\otimes X_u^{\otimes(\ell-i)}\Big]\Pi_\ell ,
\]
and $\frac{d}{du}\log Z(X_u)=\Tr((I\pm X_u)^{-1}Y)$ by Jacobi's formula applied to $\log Z=\pm\log\det(I\pm X)$, the two signs cancelling. Split $Y=Y_+-Y_-$ into its positive and negative parts, $|Y|=Y_++Y_-$. For $W\succeq0$ each operator $\Pi_\ell(X_u^{\otimes(i-1)}\otimes W\otimes X_u^{\otimes(\ell-i)})\Pi_\ell$ is positive, being a compression of a positive operator, so its trace norm is its trace; hence
\begin{align*}
\sum_\ell\|\Gamma_\ell'(u)\|_1&\le\sum_\ell\sum_{i=1}^\ell\Tr\,\Pi_\ell\big(X_u^{\otimes(i-1)}\otimes|Y|\otimes X_u^{\otimes(\ell-i)}\big)\Pi_\ell\\
&=\frac{d}{dv}Z(X_u+v|Y|)\Big|_{v=0}=Z(X_u)\Tr\big((I\pm X_u)^{-1}|Y|\big).
\end{align*}
The middle equality holds because $\Tr\Pi_\ell(X_u+v|Y|)^{\otimes\ell}\Pi_\ell$ is a polynomial in $v$ with nonnegative coefficients whose linear coefficient is the inner sum, and $\sum_\ell$ of these polynomials is $Z(X_u+v|Y|)$, finite for $v$ small; Tonelli's theorem for the double series with nonnegative terms identifies the coefficients of the sum with the sums of the coefficients, and since the sum is finite at some $v_0>0$, namely for $v_0<(1-\|X_u\|_\op)/\|Y\|_\op$, the resulting power series converges on $|v|<v_0$ and its derivative at $0$ is its linear coefficient. (For fermions the sums over $\ell$ are finite.) Therefore
\[
\sum_\ell\|\rho_\ell'(u)\|_1\le\Tr\big((I\pm X_u)^{-1}|Y|\big)+\big|\Tr((I\pm X_u)^{-1}Y)\big|\le2\Tr\big((I\pm X_u)^{-1}|Y|\big),
\]
which is at most $2\|(I\pm X_u)^{-1}\|_\op\|Y\|_1$,
and integrating over $u$ gives the display. For fermions $\|(I+X_u)^{-1}\|_\op\le1$; for bosons $\|(I-X_u)^{-1}\|_\op\le1/\delta'$ when $\|X_u\|_\op\le1-\delta'$, which holds on the segment by convexity. The vacuum is $\rho_0$, and $\|X\|_1=s$.
\end{proof}

The fermionic case of (ii) also follows from the trace-norm perturbation bound of Bittel, Mele, Eisert and Leone \cite[Thm.~1]{BMEL25}, $\|\rho-\sigma\|_1\le\tfrac12\|\Gamma(\rho)-\Gamma(\sigma)\|_1$ for all fermionic Gaussian states, together with the resolvent identity $\|X(I+X)^{-1}-X'(I+X')^{-1}\|_1=\|(I+X)^{-1}(X-X')(I+X')^{-1}\|_1\le\|X-X'\|_1$ for the occupation matrices and $\|\Gamma-\Gamma'\|_1\le4\|\mathsf N-\mathsf N'\|_1$; the proof above is included because it covers bosons in the same three lines. The constant $2$ is sharp but not attained: for a single fermionic mode the ratio $\|\rho_x-\rho_{x'}\|_1/\|x-x'\|_1$ equals $2/((1+x)(1+x'))$, which tends to $2$ as $x,x'\to0$. Note the contrast with the Frobenius norm: no dimension-free bound $\|\rho_\Gamma-\rho_{\Gamma'}\|_1\le K\|\Gamma-\Gamma'\|_F$ holds on the full fermionic class \cite[Prop.~7.4]{Rub26}, and it is the trace norm, of the covariance or of the fugacity, that is dimension-free.

\begin{theorem}[$k$-copy rate of the cold number-conserving classes]\label{cor:kcopy-rate}
Fix $C\ge\tfrac12$ for fermions; $C\ge\tfrac13$ and $0<\delta\le\tfrac34$ for bosons. Let $m\ge\max\{m_0,2\}$ and $0<\eps<\eps_0$ with the constants of Theorem~\ref{thm:vacuum} (fermions) or of \cite[Thm.~2]{CMMP26} (bosons). For every $k\ge1$, the number of copies needed by $k$-copy protocols to learn every state of $\mathcal N^{\rm F}(C)$, respectively $\mathcal N^{\rm B}(C,\delta)$, to trace distance $\eps$ with probability $2/3$ is
\[
\Theta\Big(\frac{m^3}{\eps^2\min\{\sqrt k,m\}}\Big),
\]
with implied constants depending only on $C$, and on $\delta$ for bosons. In particular, for every constant $c\ge\tfrac23$ and $m\ge\max\{m_0,2c\}$, the passive bosonic Gaussian states with $\Sigma\preceq(\tfrac12+c/m)I$, the passive part of the cold promise class of \cite[Thm.~2]{CMMP26}, have $k$-copy rate $\Theta(m^3/(\eps^2\min\{\sqrt k,m\}))$ for all $k$; for $c<\tfrac23$ the upper bound still holds. The protocol attaining the upper bound is explicit up to the constant of \cite[Thm.~1.2]{NZ26}: a particle-number measurement on every copy, followed by the joint measurement of \cite{NZ26} on groups of one-particle copies within each block, the group size being chosen from a first single-copy stage. On a state of fugacity trace $s=\tr X$ it uses, with probability at least $0.95$,
\[
O\Big(\frac1{\eps^2}+\frac{m^3\,s}{\eps^2\max\{1,\min\{\sqrt{sk},m\}\}}\Big)
\]
copies, which is $O(m^3/(\eps^2\min\{\sqrt k,m\}))$ uniformly in $s\le C$.
\end{theorem}

\begin{proof}
\emph{Lower bound.} $\mathcal N^{\rm F}(C)\supset\{\rho^{\rm F}_w:w\preceq2I/m\}$, on which Theorem~\ref{thm:vacuum} holds for all $k$; and $\mathcal N^{\rm B}(C,\delta)\supset\{\rho_\sigma:\sigma\preceq2I/m\}$, the hard family of \cite[Thm.~2]{CMMP26} (their Section~4), or equally the bosonic form of Proposition~\ref{prop:transfer}. A lower bound on a subclass is a lower bound on the class. The class $\{\Sigma\preceq(\tfrac12+c/m)I\}$ contains the hard family when $c\ge\tfrac23$ and $m\ge2$: the fugacity eigenvalues of the hard family are at most $1/(2m)$, so each normal mode holds at most $(1/(2m))/(1-1/(2m))=1/(2m-1)$ photons, and $1/(2m-1)\le2/(3m)\le c/m$ for $m\ge2$. (The total photon number of the hard family is at most $\tfrac13$, but the promise is on each mode.)

\emph{Upper bound.} Write $\rho_X$ for the unknown state, $s=\tr X\le C$, $\tau=X/s$, and $a=P_1\ge s/Z_0$ for its one-particle weight (Lemma~\ref{lem:coldclass}(i)). All measurements below act on at most $k$ fresh copies and keep only classical information, so the protocol is in the block model of Section~\ref{sec:setting}.

\emph{Stage 1: the fugacity trace.} Measure the particle number on $N_1=\lceil C_1L^2Z_0^2(1+C)^2/\eps^2\rceil$ single copies and let $f_0,f_1$ be the observed frequencies of the outcomes $0$ and $1$. With $\eta=\eps/(8LZ_0(1+C))$ and $C_1$ an absolute constant, Hoeffding's inequality gives $|f_i-P_i|\le\eta$ for $i=0,1$ with probability at least $0.95$. On that event $f_0\ge P_0-\eta\ge1/(2Z_0)$, since $\eta\le1/(2Z_0)$ for $\eps\le1$, and
\[
\Big|\frac{f_1}{f_0}-\frac{P_1}{P_0}\Big|\le\frac{|f_1-P_1|+s\,|f_0-P_0|}{f_0}\le2Z_0(1+C)\eta=\frac\eps{4L}.
\]
If $f_0=0$, output the vacuum and stop. Put $\hat s=\min\{C,\max\{0,f_1/f_0\}\}$; clipping to $[0,C]\ni s$ does not increase the error, so $|\hat s-s|\le\eps/(4L)$. If $\hat s\le3\eps/(4L)$, output the vacuum and stop: then $s\le\eps/L$ and $T(\rho_X,|0\rangle\langle0|)\le Ls/2\le\eps/2$ by Lemma~\ref{lem:coldclass}(ii). Otherwise $\eps/(4L)<\hat s/3$, so
\[
\tfrac23\hat s<s<\tfrac43\hat s,\qquad a\ge\frac s{Z_0}>\frac{2\hat s}{3Z_0}.
\]

\emph{Stage 2: the one-particle state.} Set
\[
t=\max\Big\{1,\ \min\Big\{\Big\lfloor\frac{\hat sk}{6Z_0}\Big\rfloor,\ m^2\Big\}\Big\},\qquad \eps_1=\frac{3\eps}{16L\hat s},
\]
and let $B=\lceil C_2m^3/(\eps_1^2t^{3/2})\rceil$, where $C_2$ is such that the nonadaptive protocol of \cite[Thm.~1.2]{NZ26} with $d=r=m$ and blocks of exactly $t$ copies, applied to $B$ groups, returns a density matrix $\hat\tau$ with $\|\hat\tau-\tau\|_1\le2\eps_1$, that is $T(\hat\tau,\tau)\le\eps_1$, with probability at least $2/3$; $C_2$ is absolute, their Eq.~(79) with their $s$ equal to our $t\le m^2$, and their hypothesis on the accuracy holds since $2\eps_1<\tfrac12$. Repeat that protocol $r=54$ times on fresh groups and return the first estimate $\hat\tau$, in a fixed order, that is within trace distance $2\eps_1$ of at least $27$ of the other $53$; if none qualifies, output the vacuum and stop. By Hoeffding, at least $28>r/2$ repetitions succeed except with probability $e^{-3}<0.05$; any two successes are within $2\eps_1$ of each other, so a success qualifies, and the chosen $\hat\tau$ together with its $27$ neighbours forms a set of $28$ estimates that must meet the set of successes, whence $T(\hat\tau,\tau)\le3\eps_1$. So $rB$ groups of exactly $t$ copies of $\tau$ suffice. Take $N_2=\lceil\max\{6Z_0t(rB+1)/\hat s,\ 36Z_0/\hat s\}\rceil$ further copies in $\lceil N_2/k\rceil$ blocks of at most $k$, so that $N_2=O(Z_0trB/\hat s+Z_0/\hat s)$; if fewer than $rB$ groups are formed below, output the vacuum and stop. Since Stage~2 runs only when $\hat s>3\eps/(4L)$, and $\hat s\le C$, the total $N_1+N_2$ is bounded by a function of $(m,k,\eps,C,\delta)$ alone, so the protocol has a deterministic copy budget; the finer count in the statement holds with probability $0.95$. On each block measure the particle number on every copy; the copies are in a product state, so the outcomes are independent, each equal to $1$ with probability $a$, and each copy with outcome $1$ is left exactly in the state $\tau$ on $\Fock_1=\C^m$. Partition those copies into groups of $t$, apply the joint measurement of \cite{NZ26} to each group, record the outcomes and discard the rest; the particle-number step followed by joint measurements on disjoint subsystems is one POVM on the block. Let $Y\sim\operatorname{Bin}(N_2,a)$ be the total number of one-particle copies. If $t>1$ then $t\le\hat sk/(6Z_0)<ka/4$, so the grouping loses fewer than $t$ copies per block, at most $\lceil N_2/k\rceil\,t\le N_2a/4+t$ in all; if $t=1$ nothing is lost. By the Chernoff bound $Y\ge N_2a/2$ except with probability $e^{-N_2a/8}\le e^{-3}<0.05$, since $N_2a\ge24$; on that event the number of groups is at least $(N_2a/2-N_2a/4-t)/t=N_2a/(4t)-1\ge N_2\hat s/(6Z_0t)-1\ge rB$. Outcomes of distinct groups are independent and identically distributed, so the first $rB$ of them are exactly what the repeated qudit protocol consumes.

\emph{Output.} Let $\hat X=\hat s\hat\tau\succeq0$. For fermions output $\rho_{\hat X}$. For bosons output $\rho_{\hat X}$ if $\|\hat X\|_\op<1$ and the vacuum otherwise. On the three good events, of total failure probability at most $0.15$,
\[
\|\hat X-X\|_1\le|\hat s-s|+s\|\hat\tau-\tau\|_1\le\frac\eps{4L}+\frac43\hat s\cdot6\eps_1=\frac\eps{4L}+\frac{3\eps}{2L}\le\frac{2\eps}L .
\]
For fermions Lemma~\ref{lem:coldclass}(ii) gives $T(\rho_{\hat X},\rho_X)\le\|\hat X-X\|_1\le2\eps/L=\eps$. For bosons, $2\eps/L=\eps\delta/2\le\delta/2$, so $\|\hat X\|_\op\le\|X\|_\op+\|\hat X-X\|_\op\le1-\delta/2$, the output is $\rho_{\hat X}$, and the segment from $X$ to $\hat X$ has operator norm at most $1-\delta/2$; Lemma~\ref{lem:coldclass}(ii) with $\delta'=\delta/2$ gives $T(\rho_{\hat X},\rho_X)\le(2/\delta)\|\hat X-X\|_1\le(L/2)(2\eps/L)=\eps$.

\emph{Count.} $N_1=O(L^2Z_0^2(1+C)^2/\eps^2)$. For $N_2$, $trB/\hat s=O(m^3/(\hat s\eps_1^2\sqrt t))=O(L^2m^3\hat s/(\eps^2\sqrt t))$. If $\hat sk<12Z_0$ then $t=1$ and $N_2=O(Z_0L^2m^3\hat s/\eps^2)=O(Z_0^2L^2m^3/(k\eps^2))$. If $12Z_0\le\hat sk\le6Z_0m^2$ then $t\ge\hat sk/(12Z_0)$ and $N_2=O(Z_0^{3/2}L^2m^3\sqrt{\hat s/k}/\eps^2)$. If $\hat sk>6Z_0m^2$ then $t=m^2$ and $N_2=O(Z_0L^2m^2\hat s/\eps^2)$. Since $\hat s\asymp s$ on the Stage~1 event, in every case $N_1+N_2=O(m^3s/(\eps^2\max\{1,\min\{\sqrt{sk},m\}\})+1/\eps^2)$ with constants depending only on $C$ and $\delta$, and each case is $O(m^3/(\eps^2\min\{\sqrt k,m\}))$: in the first, $m^3/k\le m^3/\sqrt k$ and $m^3/k\le m\le m^2$ when $k\ge m^2$; in the second, $\sqrt{\hat s/k}\le\sqrt{C/k}$ and, when $k\ge m^2$, $\hat sk\le6Z_0m^2$ gives $m^3\sqrt{\hat s/k}\le\sqrt{6Z_0}\,m^4/k\le\sqrt{6Z_0}\,m^2$; in the third, $m^2\hat s\le Cm^2$, and the third case has $k>6Z_0m^2/C\ge m^2$, so $\min\{\sqrt k,m\}=m$ there. The term $Z_0/\hat s=O(LZ_0/\eps)$ is absorbed.
\end{proof}

\begin{remark}[what this settles and what it does not]\label{rem:kgap}
The outlook of \cite{CMMP26} lists ``closing the $k$-entangled gap'' first: a tight $k$-copy characterisation for Gaussian-state learning, of which they prove the lower bound $\Omega(m^3/(\eps^2\min\{\sqrt k,m\}))$ under the promise that the state is passive and cold, $\Sigma\preceq(\tfrac12+O(1/m))I$. Theorem~\ref{cor:kcopy-rate} closes it on that promise class, for every $k$ and every value at least $\tfrac23$ of the implied constant (for smaller values the upper bound holds, and the lower bound would need the local form of their Theorem~2 on a rescaled ball), and gives the fermionic statement on the number-conserving classes $\mathcal N^{\rm F}(C)$, $C\ge\tfrac12$. Cold Gaussian states that are not passive, with an unknown displacement or a small squeezing, are treated in Section~\ref{sec:cold-general}: a heterodyne stage estimates and undoes both before the readout, and Theorem~\ref{thm:cold-general} gives the same rate on the whole cold class. What remains open is the $k$-copy rate at intermediate temperatures $1/m\ll\nu\ll1$ (Question~\ref{q:upper}). For single copies the state-dependent count is $O(\eps^{-2}+m^3s/\eps^2)$, the second-stage term linear in the fugacity trace $s$, from $m^3/\eps^2$ at $s\asymp1$ down to the trivial regime $s\lesssim\eps$ where the vacuum is an $\eps$-accurate output; with blocks of $k$ the second-stage term is $m^3s/(\eps^2\max\{1,\min\{\sqrt{sk},m\}\})$, so $\sqrt{s/k}$ in the unsaturated block regime $1\lesssim sk\lesssim m^2$. A matching lower bound $\Omega(m^3s/\eps^2)$ for single copies is what the sampler of Section~\ref{sec:vacuum} with the rescaled law $\Prob(L=\ell)\propto s^\ell$ would be expected to give, for $\eps\ll s$; it is not proved here. Finally, the near-vacuum bosonic class $\mathcal Q_m$ of \cite[Sec.~9]{Rub26}, fugacity $w/2$, has at most one photon in expectation and fugacity eigenvalues at most $\tfrac12$, so lies in $\mathcal N^{\rm B}(1,\tfrac12)$, and its $k$-copy rate is also determined, the lower bound coming from Proposition~\ref{prop:transfer} in its bosonic form rather than from \cite[Thm.~2]{CMMP26} directly.
\end{remark}

\subsection{The cold bosonic class, displacement and squeezing included}\label{sec:cold-general}

Theorem~\ref{cor:kcopy-rate} needs the state to be passive because the readout is the one-particle block. A cold Gaussian state that is not passive differs from a passive one by a displacement and by a squeezing of order $1/m$ per mode, and both can be estimated by heterodyne detection and undone before the readout. Two facts make this cheap. The residual displacement and squeezing left after the correction enter the trace distance through a \emph{Frobenius} norm, the norm in which a $2m\times2m$ covariance matrix is learned by heterodyne from $m^2/\eps^2$ copies, and not through the trace norm that governs the thermal part and costs the factor $m$; and the squeezing part of a covariance matrix is an explicit Lipschitz function of the covariance near the vacuum, even though the Williamson normal form, whose symplectic eigenvalues are all nearly degenerate there, is not.

\paragraph{Notation.} On $m$ bosonic modes let $R=(q_1,\dots,q_m,p_1,\dots,p_m)$, $[R_j,R_k]=i\Omega_{jk}$ with $\Omega=\begin{psmallmatrix}0&I\\-I&0\end{psmallmatrix}$, so $\Omega^T=\Omega^{-1}=-\Omega$. A state with finite second moments has mean $\mu_j=\Tr(R_j\rho)$ and covariance $\Sigma_{jk}=\tfrac12\Tr(\{R_j-\mu_j,R_k-\mu_k\}\rho)$, with $\Sigma+\tfrac i2\Omega\succeq0$; the vacuum has $\Sigma=\tfrac12I$, and a Gaussian state $\rho(\mu,\Sigma)$ is determined by $(\mu,\Sigma)$ \cite[Sec.~2]{CMMP26}. A real $2m\times2m$ matrix $S$ is symplectic if $S\Omega S^T=\Omega$; the Gaussian unitary $U_S$ with $U_S^\dagger RU_S=SR$ maps $\rho(\mu,\Sigma)$ to $\rho(S\mu,S\Sigma S^T)$, and the displacement $D(v)=\exp(-iv^T\Omega R)$ maps it to $\rho(\mu+v,\Sigma)$. A symplectic matrix that is also orthogonal is the realification of a unitary $U\in\mathrm U(m)$ and $U_S=\Gamma(U)$ is passive; a Gaussian state is passive if and only if $\mu=0$ and $\Sigma$ commutes with $\Omega$, in which case $\Sigma=\Phi_{\rm re}(\mathsf N+\tfrac12I)$ for its occupation matrix $\mathsf N$ \cite[Sec.~4]{CMMP26} and the state is $\rho_X$ of Section~\ref{sec:kcopy-rate} with $X=\mathsf N(I+\mathsf N)^{-1}$. Every symmetric matrix $E$ splits as $E=E_c+E_a$ with $E_c=\tfrac12(E-\Omega E\Omega)$ commuting and $E_a=\tfrac12(E+\Omega E\Omega)$ anticommuting with $\Omega$; a symmetric $G$ anticommuting with $\Omega$ is a Hamiltonian matrix, $G\Omega+\Omega G=0$, so $P=e^G$ is symplectic, symmetric and positive, a pure multimode squeezer; such a $G$ has spectrum symmetric about $0$ and $\Tr G=0$, and $\Tr(GE_c)=0$ for every $E_c$ commuting with $\Omega$, since $GE_c$ anticommutes with $\Omega$. Norms: $\|\cdot\|_\op$, $\|\cdot\|_F$ and $\|\cdot\|_1$ on $2m\times2m$ real matrices.

\begin{lemma}[the squeezing factor]\label{lem:squeezer}
Let $\Sigma$ be the covariance of an $m$-mode Gaussian state, $\Sigma=SDS^T$ its Williamson form with $S$ symplectic and $D=\operatorname{diag}(d,d)$, $d_j\ge\tfrac12$.
\begin{enumerate}
\item[(i)] $P:=(SS^T)^{1/2}$ is symmetric, positive and symplectic, $P=e^G$ with $G$ symmetric and anticommuting with $\Omega$, $\Sigma=P\Sigma_pP$ with $\Sigma_p=P^{-1}\Sigma P^{-1}$ passive, and $P$ is a function of $\Sigma$ alone:
\[
Q:=\big(-(\Sigma\Omega)^2\big)^{1/2}=SDS^{-1},\qquad P^2=Q^{-1}\Sigma ,
\]
the square root being the principal one. Hence $\rho(\mu,\Sigma)=D(\mu)U_P\rho_pU_P^\dagger D(\mu)^\dagger$ with $\rho_p$ the passive Gaussian state of covariance $\Sigma_p$. Moreover $\|P\|_\op^2=e^{2\|G\|_\op}\le2\|\Sigma\|_\op$.
\item[(ii)] If $\|\Sigma\|_\op\le\tfrac{11}{20}$ then $\|G\|_F\le2\|(\Sigma-\tfrac12I)_a\|_F$.
\item[(iii)] If $\Sigma,\tilde\Sigma$ are covariances of Gaussian states with $\|\Sigma\|_\op,\|\tilde\Sigma\|_\op\le\tfrac{11}{20}$ and squeezing factors $P,\tilde P$, then $\|P-\tilde P\|_F\le3\|\Sigma-\tilde\Sigma\|_F$.
\item[(iv)] For $G$ as in (i) and any symmetric $\Sigma$ with $E=\Sigma-\tfrac12I$, $\Tr(e^{-2G}\Sigma)\le\Tr\Sigma+2\|G\|_F\|E_a\|_F+2e^{2\|G\|_\op}\|\Sigma\|_\op\|G\|_F^2$.
\end{enumerate}
\end{lemma}

\begin{proof}
(i) $SS^T$ is symmetric positive and symplectic, and the real powers of a positive symplectic matrix are symplectic, so $P$ is symplectic; $\log P=G$ is symmetric and, since $P\Omega P^T=\Omega$ gives $\Omega P\Omega^{-1}=P^{-1}$, $\Omega G\Omega^{-1}=-G$: $G$ anticommutes with $\Omega$. Its spectrum is symmetric about $0$, so $\|P\|_\op=\|P^{-1}\|_\op=e^{\|G\|_\op}$. Put $K=P^{-1}S$; then $KK^T=P^{-1}SS^TP^{-1}=I$ and $K$ is symplectic, so $\Sigma_p:=KDK^T=P^{-1}\Sigma P^{-1}$ is passive and $\Sigma=P\Sigma_pP$. From $S^T\Omega=\Omega S^{-1}$ and $D\Omega D\Omega=D^2\Omega^2=-D^2$ we get $\Sigma\Omega=SD\Omega S^{-1}$ and $-(\Sigma\Omega)^2=SD^2S^{-1}$, a diagonalisable matrix with spectrum $\{d_j^2\}\subset[\tfrac14,\infty)$, whose principal square root is $Q=SDS^{-1}$; then $Q^{-1}\Sigma=SD^{-1}S^{-1}SDS^T=SS^T=P^2$. The factorisation of $\rho(\mu,\Sigma)$ follows from the action of $D(\mu)$ and $U_P$ on $(\mu,\Sigma)$. Finally $\Sigma=P\Sigma_pP\succeq\tfrac12P^2$ since $\Sigma_p\succeq\tfrac12I$, whence $e^{2\|G\|_\op}=\|P^2\|_\op\le2\|\Sigma\|_\op$.

(ii) Write $g=\|G\|_\op$, $\mathrm{Ch}=\cosh G$, $\mathrm{Sh}=\sinh G$ (even and odd functions of $G$, so $\mathrm{Ch}$ commutes and $\mathrm{Sh}$ anticommutes with $\Omega$), and $E_p=\Sigma_p-\tfrac12I$, which commutes with $\Omega$ and satisfies $0\preceq E_p$, $\|E_p\|_\op\le\|P^{-1}\|_\op^2\|\Sigma\|_\op-\tfrac12\le2\|\Sigma\|_\op^2-\tfrac12$. Then $\Sigma=\tfrac12e^{2G}+(\mathrm{Ch}+\mathrm{Sh})E_p(\mathrm{Ch}+\mathrm{Sh})$, and the anticommuting part of $\Sigma-\tfrac12I$ is
\[
(\Sigma-\tfrac12I)_a=\tfrac12\sinh(2G)+\mathrm{Ch}\,E_p\,\mathrm{Sh}+\mathrm{Sh}\,E_p\,\mathrm{Ch} .
\]
Since $\tfrac12\sinh(2x)\ge x$ for $x\ge0$ and both sides are odd, $\|\tfrac12\sinh(2G)\|_F\ge\|G\|_F$; and $\|\mathrm{Ch}E_p\mathrm{Sh}+\mathrm{Sh}E_p\mathrm{Ch}\|_F\le2\cosh(g)\|E_p\|_\op\|\sinh G\|_F\le2\cosh(g)\frac{\sinh g}{g}\|E_p\|_\op\|G\|_F=\frac{\sinh 2g}{g}\|E_p\|_\op\|G\|_F$. For $\|\Sigma\|_\op\le\tfrac{11}{20}$ we have $e^{2g}\le\tfrac{11}{10}$, $g\le0.048$, $\sinh(2g)/g\le2\cosh(2g)\le2.01$ and $\|E_p\|_\op\le0.105$, so the subtracted term is at most $0.22\|G\|_F$ and $\|G\|_F\le\|(\Sigma-\tfrac12I)_a\|_F/0.78\le2\|(\Sigma-\tfrac12I)_a\|_F$.

(iii) Put $X=-(\Sigma\Omega)^2=SD^2S^{-1}$, $\tilde X$ likewise, and $\kappa=\|P\|_\op^2\le\tfrac{11}{10}$, $\tilde\kappa$ likewise. For $t\ge0$, $\|(X+tI)^{-1}\|_\op=\|S(D^2+tI)^{-1}S^{-1}\|_\op\le\kappa/(\tfrac14+t)$. The principal square root has the representation $\sqrt X=\frac1\pi\int_0^\infty t^{-1/2}X(X+tI)^{-1}\,dt$, valid on each eigenvalue, and $X(X+t)^{-1}-\tilde X(\tilde X+t)^{-1}=t(\tilde X+t)^{-1}(X-\tilde X)(X+t)^{-1}$, so
\[
\|Q-\tilde Q\|_F\le\frac{\kappa\tilde\kappa}{\pi}\int_0^\infty\frac{t^{1/2}\,dt}{(\tfrac14+t)^2}\,\|X-\tilde X\|_F=\kappa\tilde\kappa\|X-\tilde X\|_F\le\kappa\tilde\kappa(\|\Sigma\|_\op+\|\tilde\Sigma\|_\op)\|\Sigma-\tilde\Sigma\|_F ,
\]
using $X-\tilde X=-(\Sigma-\tilde\Sigma)\Omega\Sigma\Omega-\tilde\Sigma\Omega(\Sigma-\tilde\Sigma)\Omega$. Next $P^2-\tilde P^2=Q^{-1}(\Sigma-\tilde\Sigma)+Q^{-1}(\tilde Q-Q)\tilde P^2$ with $\|Q^{-1}\|_\op=\|SD^{-1}S^{-1}\|_\op\le2\kappa$, so $\|P^2-\tilde P^2\|_F\le(2\kappa+2\kappa^2\tilde\kappa^2(\|\Sigma\|_\op+\|\tilde\Sigma\|_\op))\|\Sigma-\tilde\Sigma\|_F\le5.5\,\|\Sigma-\tilde\Sigma\|_F$. Finally, for positive definite $A,B$ the matrix $Y=\sqrt A-\sqrt B$ solves $\sqrt A\,Y+Y\sqrt B=A-B$, whose solution in the eigenbases of $\sqrt A$ and $\sqrt B$ is entrywise $(A-B)_{jk}/(a_j+b_k)$, so $\|\sqrt A-\sqrt B\|_F\le\|A-B\|_F/(\lambda_{\min}(\sqrt A)+\lambda_{\min}(\sqrt B))$; with $\lambda_{\min}(P)=\kappa^{-1/2}\ge(10/11)^{1/2}$ this gives $\|P-\tilde P\|_F\le5.5\cdot\tfrac12(11/10)^{1/2}\|\Sigma-\tilde\Sigma\|_F\le3\|\Sigma-\tilde\Sigma\|_F$.

(iv) $\Tr(e^{-2G}\Sigma)=\Tr\Sigma-2\Tr(G\Sigma)+\Tr((e^{-2G}-I+2G)\Sigma)$. In the first correction $\Tr(G\Sigma)=\tfrac12\Tr G+\Tr(GE_c)+\Tr(GE_a)=\Tr(GE_a)$, of modulus at most $\|G\|_F\|E_a\|_F$. In the second, $e^{-2G}-I+2G=f(G)$ with $f(x)=e^{-2x}-1+2x\in[0,2x^2e^{2|x|}]$, so $\|f(G)\|_1\le2e^{2\|G\|_\op}\|G\|_F^2$ and $|\Tr(f(G)\Sigma)|\le\|f(G)\|_1\|\Sigma\|_\op$.
\end{proof}

\begin{lemma}[trace distance under a displacement or a squeezing]\label{lem:perturb}
Let $\rho$ be a state and $H$ a self-adjoint operator such that $\rho=\sum_jp_j|\psi_j\rangle\langle\psi_j|$ with every $\psi_j$ in the domain of $H$ and $\operatorname{Var}_\rho(H):=\Tr(H^2\rho)-(\Tr H\rho)^2<\infty$. Then
\[
T\big(\rho,\ e^{-iH}\rho\,e^{iH}\big)\le\sqrt{\operatorname{Var}_\rho(H)} .
\]
For a Gaussian state $\rho(\mu,\Sigma)$ and $v\in\R^{2m}$ this gives $T(\rho,D(v)\rho D(v)^\dagger)\le\|\Sigma\|_\op^{1/2}\|v\|_2$; and for a zero-mean Gaussian state $\rho(0,\Sigma)$ and a symmetric $G$ anticommuting with $\Omega$,
\[
T\big(\rho(0,\Sigma),\,U_{e^G}\rho(0,\Sigma)\,U_{e^G}^\dagger\big)\le\Big(\tfrac12\|\Sigma\|_\op^2+\tfrac18\Big)^{1/2}\|G\|_F .
\]
(For a mean $\mu\ne0$ the variance of the quadratic generator acquires the term $(h\mu)^T\Sigma(h\mu)$ with $h=\Omega^TG$, since the squeezer moves the mean to $e^G\mu$; the bound is not uniform in $\mu$, and is used below only at $\mu=0$.)
\end{lemma}

\begin{proof}
Let $c=\Tr(H\rho)$ and $U=e^{-iH}$. Trace distance is jointly convex, so $T(\rho,U\rho U^\dagger)\le\sum_jp_jT(\psi_j,U\psi_j)$. For a unit vector $\psi$ and the unit vector $\phi=e^{ic}U\psi=e^{-i(H-c)}\psi$, $T(\psi,\phi)=(1-|\langle\psi,\phi\rangle|^2)^{1/2}\le\|\phi-\psi\|$, because $|\langle\psi,\phi\rangle|\ge\operatorname{Re}\langle\psi,\phi\rangle=1-\tfrac12\|\phi-\psi\|^2$; and $\|\phi-\psi\|=\|(e^{-i(H-c)}-I)\psi\|\le\|(H-c)\psi\|$ by the spectral theorem, since $|e^{-ix}-1|\le|x|$. Hence $T(\rho,U\rho U^\dagger)\le\sum_jp_j\|(H-c)\psi_j\|\le(\sum_jp_j\|(H-c)\psi_j\|^2)^{1/2}=\operatorname{Var}_\rho(H)^{1/2}$ by Cauchy--Schwarz. A Gaussian state has finite moments of all orders and its eigenvectors, Gaussian-unitary images of Fock states, lie in the domain of every polynomial in $R$. For $D(v)=e^{-iH}$ with $H=v^T\Omega R$, linear in $R$, $\operatorname{Var}_\rho(H)=(\Omega v)^T\Sigma(\Omega v)\le\|\Sigma\|_\op\|v\|_2^2$. For $U_{e^G}=e^{-iH}$ with $H=\tfrac12R^ThR$, $h=\Omega^TG$, which is symmetric since $(\Omega^TG)^T=G\Omega=-\Omega G=\Omega^TG$, the Heisenberg equation $\tfrac{d}{du}e^{iuH}Re^{-iuH}=\Omega h\,R$ gives the symplectic matrix $e^{\Omega h}=e^{G}$, and $\|h\|_F=\|G\|_F$. For the zero-mean state, by Wick's theorem, with $W=\Sigma+\tfrac i2\Omega$ the matrix of $\Tr(R_jR_k\rho)$, $\Tr(R_iR_jR_kR_l\rho)=W_{ij}W_{kl}+W_{ik}W_{jl}+W_{il}W_{jk}$, and since $h$ is symmetric the last two pairings contribute $\Tr(hWhW^T)$ each, so
\[
\operatorname{Var}_\rho(\tfrac12R^ThR)=\tfrac12\Tr(hWhW^T)=\tfrac12\Tr\big((h\Sigma)^2\big)+\tfrac18\Tr\big((h\Omega)^2\big),
\]
the cross terms cancelling by cyclicity. (On the vacuum with $h=I$ this is $0$, as it must be for $H=\sum_j(a_j^\dagger a_j+\tfrac12)$.) Now $\Tr((h\Sigma)^2)\le\|h\|_F^2\|\Sigma\|_\op^2$ and $\Tr((h\Omega)^2)\le\|h\Omega\|_F^2=\|h\|_F^2$.
\end{proof}

For $c>0$ let
\[
\mathcal G^{\rm B}_m(c)=\big\{\rho(\mu,\Sigma):\ \mu\in\R^{2m},\ \Sigma\preceq(\tfrac12+c/m)I\big\},
\]
the cold Gaussian states of \cite{CMMP26} with no restriction on the displacement; it contains the passive class $\{\Sigma\preceq(\tfrac12+c/m)I\}$ of Theorem~\ref{cor:kcopy-rate}. Its states have at most $c/m$ thermal photons per mode and squeezing at most $c/m$ per mode: by Lemma~\ref{lem:squeezer}(i), $e^{2\|G\|_\op}\le1+2c/m$.

\begin{theorem}[$k$-copy rate of the cold bosonic class]\label{thm:cold-general}
Fix $c>0$. There are $m_\star(c)$ and $\eps_\star(c)>0$ such that for $m\ge m_\star$, $0<\eps<\eps_\star$ and every $k\ge1$, a $k$-copy protocol learns every state of $\mathcal G^{\rm B}_m(c)$ to trace distance $\eps$ with probability at least $2/3$ from
\[
O\Big(\frac{m^3}{\eps^2\min\{\sqrt k,m\}}\Big)
\]
copies, with the implied constant depending only on $c$. For $c\ge\tfrac23$ this is the $k$-copy rate of $\mathcal G^{\rm B}_m(c)$, the lower bound being \cite[Thm.~2]{CMMP26}. The protocol is: heterodyne detection on $O(m^2/\eps^2)$ copies, from which the mean and the squeezing factor of Lemma~\ref{lem:squeezer}(i) are estimated; the inverse displacement and inverse squeezer applied to every further copy; the protocol of Theorem~\ref{cor:kcopy-rate} on those copies; and the output of that protocol displaced and squeezed back.
\end{theorem}

\begin{proof}
\emph{Lower bound.} As in Theorem~\ref{cor:kcopy-rate}: the passive hard family of \cite[Thm.~2]{CMMP26} lies in $\mathcal G^{\rm B}_m(c)$ for $c\ge\tfrac23$, $m\ge2$.

\emph{Upper bound.} Let $\rho=\rho(\mu,\Sigma)\in\mathcal G^{\rm B}_m(c)$, $\Sigma=P\Sigma_pP$ by Lemma~\ref{lem:squeezer}(i), and put $C_\star=c+1$, $Z_\star=2e^{C_\star}$, $K_\star=2\big(16e^{C_\star}(1+C_\star)+1\big)$. We take $m_\star\ge\max\{m_0,100c,100c^2\}$, so that $\|\Sigma\|_\op\le0.51$ and $c/\sqrt m\le\tfrac1{10}$, and $\eps_\star\le\min\{\eps_0,\tfrac1{20}\}$ small enough for the finitely many inequalities below, each of which holds for $\eps$ below a constant depending on $c$; $\eps_0$ and $m_0$ are the constants of \cite[Thm.~2]{CMMP26}, so that the lower bound holds on the same range.

\emph{Stage 0: heterodyne.} Heterodyne detection on a copy of $\rho$ returns $y\sim\mathcal N(\mu,C)$ with $C=\Sigma+\tfrac12I\preceq1.01\,I$ \cite[Sec.~2]{CMMP26}. Take $N_0=2n_0$ copies, $n_0=\lceil K_0m^2/\eps^2\rceil$, and let $\hat\mu$ be the sample mean of all $N_0$ outcomes and $\hat C=n_0^{-1}\sum_{i\le n_0}z_iz_i^T$ with $z_i=(y_{2i}-y_{2i-1})/\sqrt2\sim\mathcal N(0,C)$ independent. Then $\hat\mu-\mu\sim\mathcal N(0,C/N_0)$, so $\E\|\hat\mu-\mu\|_2^2=\Tr C/N_0\le1.01\,m/n_0$ and $\|\hat\mu-\mu\|_2\le(21m/n_0)^{1/2}$ except with probability $0.05$; and by the covariance estimate for Gaussian vectors, $\E\|\hat C-C\|_\op\le K'\big((2m/n_0)^{1/2}+2m/n_0\big)\|C\|_\op$ for an absolute constant $K'$ \cite[Thm.~4.7.1]{Ver18}, together with Markov's inequality, $\|\hat C-C\|_\op\le\eta:=K_V(m/n_0)^{1/2}$ except with probability $0.05$, with $K_V=60K'$ and $n_0\ge2m$. Put $\hat\Sigma=\hat C-\tfrac12I+\eta I$. On the good event $\hat\Sigma-\Sigma=\hat C-C+\eta I\succeq0$, so $\hat\Sigma+\tfrac i2\Omega\succeq\Sigma+\tfrac i2\Omega\succeq0$ and $\hat\Sigma$ is the covariance of a Gaussian state, with $\|\hat\Sigma\|_\op\le0.51+2\eta\le0.52$ and
\begin{align*}
\|\hat\Sigma-\Sigma\|_F&\le\sqrt{2m}\,\|\hat\Sigma-\Sigma\|_\op\le2\sqrt{2m}\,\eta=2\sqrt2K_V\,m/\sqrt{n_0}\le2\sqrt2K_V\,\eps/\sqrt{K_0},\\
\|\hat\mu-\mu\|_2&\le\sqrt{21}\,\eps/\sqrt{K_0m},
\end{align*}
and $K_0=K_0(c)$ is chosen so that $\|\hat\mu-\mu\|_2+5\|\hat\Sigma-\Sigma\|_F\le\eps/K_\star$ and $\|\hat\Sigma-\Sigma\|_F\le1/100$. If $\hat\Sigma+\tfrac i2\Omega\not\succeq0$ or $\|\hat\Sigma\|_\op>0.52$, output the vacuum and stop; on the good event neither happens. Otherwise let $\hat P$ be the squeezing factor of $\hat\Sigma$, $\hat P=e^{\hat G}$, and $V=U_{\hat P}^\dagger D(\hat\mu)^\dagger$, a Gaussian unitary.

\emph{Stage 1: undoing.} Every further copy is first transformed by $V$; by Lemma~\ref{lem:squeezer}(i) the copies are then in the Gaussian state $\rho'=V\rho V^\dagger=\rho(\mu',\Sigma')$ with $\mu'=\hat P^{-1}(\mu-\hat\mu)$ and $\Sigma'=\hat P^{-1}\Sigma\hat P^{-1}$. Write $\Sigma'=P'\Sigma_{p'}P'$, $P'=e^{G'}$, by Lemma~\ref{lem:squeezer}(i), and let $\rho_{p'}$ be the passive Gaussian state of covariance $\Sigma_{p'}$, so that $\rho'=D(\mu')U_{P'}\rho_{p'}U_{P'}^\dagger D(\mu')^\dagger$. We bound $\theta:=T(\rho',\rho_{p'})$. By Lemma~\ref{lem:squeezer}(i), $\|\hat P^{\pm1}\|_\op^2\le2\|\hat\Sigma\|_\op\le1.04$, so $\|\Sigma'\|_\op\le1.04\cdot0.51\le0.54$, $\|\Sigma_p\|_\op\le1.02\cdot0.51\le0.53$ and $\|\Sigma_{p'}\|_\op\le1.08\cdot0.54\le0.59$. Put $W=\hat P^{-1}P-I$; by Lemma~\ref{lem:squeezer}(iii), $\|W\|_F\le\|\hat P^{-1}\|_\op\|P-\hat P\|_F\le1.02\cdot3\|\hat\Sigma-\Sigma\|_F\le3.1\|\hat\Sigma-\Sigma\|_F\le0.04$. Since $\Sigma'=(I+W)\Sigma_p(I+W)^T$ and $\Sigma_p$ commutes with $\Omega$,
\[
(\Sigma'-\tfrac12I)_a=(W\Sigma_p+\Sigma_pW^T+W\Sigma_pW^T)_a,
\]
so $\|(\Sigma'-\tfrac12I)_a\|_F\le\|\Sigma_p\|_\op(2+\|W\|_\op)\|W\|_F\le4\|\hat\Sigma-\Sigma\|_F$, and Lemma~\ref{lem:squeezer}(ii), applicable since $\|\Sigma'\|_\op\le\tfrac{11}{20}$, gives $\|G'\|_F\le8\|\hat\Sigma-\Sigma\|_F$. Lemma~\ref{lem:perturb}, applied to the displacement of $U_{P'}\rho_{p'}U_{P'}^\dagger$, of covariance $\Sigma'$, and to the squeezing of $\rho_{p'}$, gives
\begin{align*}
\theta&\le\|\Sigma'\|_\op^{1/2}\|\mu'\|_2+\big(\tfrac12\|\Sigma_{p'}\|_\op^2+\tfrac18\big)^{1/2}\|G'\|_F\\
&\le0.74\cdot1.02\,\|\mu-\hat\mu\|_2+0.55\cdot8\,\|\hat\Sigma-\Sigma\|_F\le\|\mu-\hat\mu\|_2+5\|\hat\Sigma-\Sigma\|_F\le\eps/K_\star .
\end{align*}
Next, $\rho_{p'}\in\mathcal N^{\rm B}(C_\star,\tfrac12)$. Its occupation matrix $\mathsf N'$ satisfies $\mathsf N'\preceq(\|\Sigma_{p'}\|_\op-\tfrac12)I\preceq0.09\,I$, so its fugacity $X_{p'}=\mathsf N'(I+\mathsf N')^{-1}$ has $\|X_{p'}\|_\op\le0.09$; and $\Tr\mathsf N'=\tfrac12\Tr\Sigma_{p'}-\tfrac m2$ with $\Tr\Sigma_{p'}=\Tr(e^{-2G'}\Sigma')$ and $\Tr\Sigma'=\Tr(e^{-2\hat G}\Sigma)$. Lemma~\ref{lem:squeezer}(iv) bounds both. For the first, $\|(\Sigma'-\tfrac12I)_a\|_F\le4\|\hat\Sigma-\Sigma\|_F$ and $\|G'\|_F\le8\|\hat\Sigma-\Sigma\|_F$, so $\Tr\Sigma_{p'}\le\Tr\Sigma'+(64+2e^{0.1}\cdot0.54\cdot64)\|\hat\Sigma-\Sigma\|_F^2\le\Tr\Sigma'+1$ for $\eps\le\eps_\star$. For the second, $E=\Sigma-\tfrac12I$ has $\|E\|_\op\le c/m$, since $\Sigma\preceq(\tfrac12+c/m)I$ and $\Sigma\succeq\tfrac12P^2\succeq\tfrac12e^{-2\|G\|_\op}I\succeq(\tfrac12-c/m)I$, so $\|E_a\|_F\le\|E\|_F\le c\sqrt{2/m}$; and $\|\hat G\|_F\le2\|(\hat\Sigma-\tfrac12I)_a\|_F\le2\|E_a\|_F+2\|\hat\Sigma-\Sigma\|_F\le3c/\sqrt m+\eps$ by Lemma~\ref{lem:squeezer}(ii), so $\Tr\Sigma'\le\Tr\Sigma+2(3c/\sqrt m+\eps)c\sqrt{2/m}+2e^{0.1}\cdot0.51\,(3c/\sqrt m+\eps)^2\le\Tr\Sigma+0.24$ for $m\ge m_\star$, $\eps\le\eps_\star$, since then $3c/\sqrt m+\eps\le0.35$ and $c\sqrt{2/m}\le\sqrt2/10$. Since $\Tr\Sigma\le m+2c$, $\Tr\mathsf N'\le c+1=C_\star$.

\emph{Stage 2: the readout.} The proof of the upper bound of Theorem~\ref{cor:kcopy-rate} uses about the measured state only this: the particle-number outcomes on fresh copies are independent, with $\Prob(0)=a_0$ and $\Prob(1)=a_0s'$ where $a_0=\langle0|\rho'|0\rangle\ge1/Z_\star$ and $s'=\tr X'$ for the operator $X':=a_0^{-1}\Pi_1\rho'\Pi_1\succeq0$ on $\Fock_1=\C^m$, with $s'$ at most the clipping constant; and the state of a copy on outcome $1$ is exactly $X'/s'$. These hold for every state with $a_0\ge1/Z_\star$, Gaussian or not, with $s'\le1/a_0\le Z_\star$. Here $a_0\ge\langle0|\rho_{p'}|0\rangle-\theta\ge e^{-C_\star}-\eps/K_\star\ge1/Z_\star$ for $\eps\le\eps_\star$. So the Stage~1 and Stage~2 protocol of Theorem~\ref{cor:kcopy-rate}, with $Z_0$ and $C$ both replaced by $Z_\star$ (in $N_1$ and in the clipping interval $[0,Z_\star]$), $\delta=\tfrac12$ (so $L=8$) and accuracy $\eps/2$, run on copies of $\rho'$, returns $\hat X\succeq0$ with $\|\hat X-X'\|_1\le\eps/L$ except with probability $0.15$, from $O(1/\eps^2+m^3/(\eps^2\min\{\sqrt k,m\}))$ copies, the constants depending on $C_\star$ only. In this use of the subroutine a vacuum return means $\hat X=0$, and every returned estimate, $\hat X=0$ included, is transformed back by $V^\dagger$ below; on the good frequency event the small-$\hat s$ stopping branch has $s'\le\eps/(2L)$, so $\|0-X'\|_1\le\eps/L$ as required. Output $V^\dagger\rho_{\hat X}V=D(\hat\mu)U_{\hat P}\rho_{\hat X}U_{\hat P}^\dagger D(\hat\mu)^\dagger$, a Gaussian state, if $\|\hat X\|_\op<1$, and the vacuum otherwise; on the good events $\|\hat X\|_\op\le\tfrac12$, as shown next.

\emph{Error.} By unitary invariance $T(V^\dagger\rho_{\hat X}V,\rho)=T(\rho_{\hat X},\rho')\le T(\rho_{\hat X},\rho_{p'})+\theta$. The pinching $\Lambda$ onto the particle-number sectors contracts the trace norm and fixes $\rho_{p'}=\rho_{X_{p'}}$, whose vacuum weight is $b_0=1/Z(X_{p'})$ and whose one-particle block is $b_0X_{p'}$; so $|a_0-b_0|+\|a_0X'-b_0X_{p'}\|_1\le\|\Lambda(\rho')-\rho_{p'}\|_1\le2\theta$, and
\[
\|X'-X_{p'}\|_1\le\frac{\|a_0X'-b_0X_{p'}\|_1}{a_0}+\tr X_{p'}\,\frac{|b_0-a_0|}{a_0}\le2\theta Z_\star(1+C_\star)=4e^{C_\star}(1+C_\star)\,\theta ,
\]
using $\tr X_{p'}\le\Tr\mathsf N'\le C_\star$. Hence $\|\hat X-X_{p'}\|_1\le\eps/8+4e^{C_\star}(1+C_\star)\theta\le\tfrac14$, the segment from $X_{p'}$ to $\hat X$ has operator norm at most $0.09+\tfrac14\le\tfrac12$, the output $\rho_{\hat X}$ is a state, and Lemma~\ref{lem:coldclass}(ii) with $\delta'=\tfrac12$ gives $T(\rho_{\hat X},\rho_{p'})\le4\|\hat X-X_{p'}\|_1\le\eps/2+16e^{C_\star}(1+C_\star)\theta$. Altogether
\[
T(V^\dagger\rho_{\hat X}V,\rho)\le\frac\eps2+\big(16e^{C_\star}(1+C_\star)+1\big)\theta\le\frac\eps2+\frac{K_\star}{2}\cdot\frac\eps{K_\star}=\eps
\]
except with probability $0.05+0.05+0.15<\tfrac13$.

\emph{Count.} $N_0=O(m^2/\eps^2)$ and $m^2\le m^3/\min\{\sqrt k,m\}$, so the total is $O(m^3/(\eps^2\min\{\sqrt k,m\}))$ with a constant depending only on $c$.
\end{proof}

\begin{remark}\label{rem:cold-general}
Theorem~\ref{thm:cold-general} closes the first open problem of \cite{CMMP26} on the cold class, the promise class of their Theorem~2, now with any displacement and squeezing: their lower bound holds even under the promise that the state is passive and cold, $\Sigma\preceq(\tfrac12+O(1/m))I$, and the upper bound now holds on all cold states. Their question is posed for Gaussian-state learning as such, and at intermediate and high temperatures, where the single-copy rate is $\Theta(m^2\min\{m,1+\nu^{-1}\}/\eps^2)$, it is not answered here. The heterodyne stage costs $m^2/\eps^2$, the collective rate, and is never the bottleneck, because the residual displacement and squeezing after the correction enter the trace distance through Frobenius norms (Lemma~\ref{lem:perturb}), whereas the thermal part enters through the trace norm of the fugacity difference (Lemma~\ref{lem:coldclass}(ii)) and costs the extra factor $\max\{m/\sqrt k,1\}$. This is the same division as in the single-copy protocol of \cite{BMEL25,CMMP26}, where heterodyne learns the covariance to Frobenius accuracy $\eps$ at cost $m^2/\eps^2$ and the conversion to trace distance costs the factor $m$; here the conversion is done by the one-particle readout, which is where the block measurements act. The constants $m_\star$, $\eps_\star$, $K_0$ can be chosen to depend only on $c$, $K_V$ and the constant of \cite[Thm.~1.2]{NZ26}; no attempt was made to optimise them. The fermionic analogue, Gaussian states near the vacuum with pairing terms, is not covered: there is no heterodyne detection for fermions, and the single-copy estimators of the fermionic covariance that are available, the matchgate shadows, cost $\Theta(m^3/\eps^2)$ for Frobenius accuracy $\eps$ in the worst case over mixed states \cite{Rub26}; whether a cheaper Frobenius stage exists near the vacuum is not known. What remains open on the bosonic side is the $k$-copy rate at intermediate temperatures (Question~\ref{q:upper}).
\end{remark}

\section{Superselection rules}\label{sec:superselection}

Physical fermionic systems obey the parity superselection rule: the observables that can be measured commute with the parity $P$, and in a learning experiment the preparations and the measurements are even. Möbus, Stilck Fran\c ca and Rouz\'e \cite{MSR26} learn local fermionic Lindbladians under exactly this constraint, citing \cite{Rub26} for the separation between collective and single-copy measurements, and ask, as a further direction, which Hamiltonian and dissipative coefficients remain learnable when preparations, controls and measurements are in addition number-conserving. For the tomography of Gaussian states neither constraint is visible, and the separation survives both.

Let $a_j=\tfrac12(c_{2j-1}+ic_{2j})$ be the mode operators and $\hat N=\sum_{j=1}^ma_j^\dagger a_j=\tfrac m2+\tfrac12\sum_{j=1}^mB_{2j-1,2j}$ the number operator. Set
\[
\mathfrak u=\{\theta\in\R^p:[K_\theta,\hat N]=0\},\qquad\mathcal N=\{\rho_\theta:\theta\in\mathfrak u\}.
\]

\begin{lemma}[number-conserving Gaussian states]\label{lem:numberconserving}
$\mathfrak u$ is a linear subspace of $\R^p$ of dimension $m^2$; an orthonormal basis is given by the coefficient vectors of the $m^2$ quadratics
\[
B_{2j-1,2j}\ (1\le j\le m),\qquad \tfrac1{\sqrt2}\big(B_{2j-1,2k}-B_{2j,2k-1}\big),\quad \tfrac1{\sqrt2}\big(B_{2j-1,2k-1}+B_{2j,2k}\big)\ (1\le j<k\le m).
\]
$\mathcal N$ is the set of all full-rank fermionic Gaussian states that commute with $\hat N$: the Gibbs states $e^{-\beta H}/\Tr e^{-\beta H}$ of the number-conserving quadratic Hamiltonians $H=\sum_{j,k}h_{jk}a_j^\dagger a_k$, $h$ Hermitian.
\end{lemma}

\begin{proof}
From $a_j=\tfrac12(c_{2j-1}+ic_{2j})$ and the anticommutation relations, for $j<k$,
\begin{gather*}
a_j^\dagger a_k+a_k^\dagger a_j=\tfrac12\big(B_{2j-1,2k}-B_{2j,2k-1}\big),\qquad i\big(a_j^\dagger a_k-a_k^\dagger a_j\big)=\tfrac12\big(B_{2j-1,2k-1}+B_{2j,2k}\big),\\
a_j^\dagger a_j=\tfrac12\big(I+B_{2j-1,2j}\big).
\end{gather*}
Hence every $\sum_{j,k}h_{jk}a_j^\dagger a_k$ with $h$ Hermitian equals $\tfrac12(\tr h)I+K_\theta$ with $\theta$ in the span $\mathfrak u_0$ of the $m^2$ listed vectors, and conversely every $K_\theta$ with $\theta\in\mathfrak u_0$ is of this form. The listed vectors are orthogonal in $\R^p$ because they involve pairwise disjoint sets of pairs $(a,b)$, and of unit length by the normalisation, so $\dim\mathfrak u_0=m^2$, and $\mathfrak u_0\subseteq\mathfrak u$ since $[a_j^\dagger a_k,\hat N]=0$. Conversely, every quadratic $K_\theta$ is, up to a constant, $\sum_{j,k}h_{jk}a_j^\dagger a_k+\sum_{j<k}(\Delta_{jk}a_ja_k+\bar\Delta_{jk}a_k^\dagger a_j^\dagger)$ with $h$ Hermitian and $\Delta$ complex: for $j<k$, $\Delta_{jk}a_ja_k+\bar\Delta_{jk}a_k^\dagger a_j^\dagger=\tfrac12\big(\operatorname{Re}\Delta_{jk}(B_{2j-1,2k}+B_{2j,2k-1})+\operatorname{Im}\Delta_{jk}(B_{2j-1,2k-1}-B_{2j,2k})\big)$, which together with the two $h$-combinations above spans the four quadratics on the pairs $\{2j-1,2j\}\times\{2k-1,2k\}$, so the map $(h,\Delta)\mapsto K_\theta$ is a bijection onto $\R^p$ (the real dimensions $m^2+m(m-1)=p$ agree); and $[\hat N,a_ja_k]=-2a_ja_k$, $[\hat N,a_k^\dagger a_j^\dagger]=2a_k^\dagger a_j^\dagger$, $[\hat N,a_j^\dagger a_k]=0$, so $[K_\theta,\hat N]=0$ forces $\Delta=0$ and $\mathfrak u=\mathfrak u_0$. Finally, every $\rho_\theta$ with $\theta\in\mathfrak u$ is a full-rank Gaussian state commuting with $\hat N$; and if $\rho$ is a full-rank Gaussian state, then by the normal form \cite[Lemma~2.1 and Eq.~(2.3)]{Rub26} it is $\rho_\theta$ for the $\theta$ whose normal-form values are $\operatorname{artanh}$ of those of its covariance matrix, and if moreover $\rho$ commutes with $\hat N$, so does $K_\theta=\log(dz_\theta\rho_\theta)$, a function of $\rho$; thus $\theta\in\mathfrak u$.
\end{proof}

A measurement on $t$ copies and ancillas is \emph{parity-even} if every effect commutes with $P^{\otimes t}\otimes I$, where $P$ is the parity of one copy, and \emph{number-conserving} if every effect commutes with $\hat N_t\otimes I$, $\hat N_t=\sum_{c=1}^t\hat N^{(c)}$; a protocol is parity-even or number-conserving if every measurement it makes is.

\begin{proposition}[superselection]\label{prop:superselection}
\begin{enumerate}
\item[(i)] Let $\mathcal S$ be any class of even states; every Gaussian state is even. Every protocol of the three kinds in Table~\ref{tab:rates} (single-copy, $k$-copy, collective) can be replaced by a parity-even one with the same outcome law, hence the same error, on every state of $\mathcal S$. If every state of $\mathcal S$ commutes with $\hat N$, the same holds with number-conserving in place of parity-even. Consequently no rate in the fermionic rows of Table~\ref{tab:rates} changes when the learner is restricted to parity-even protocols, and no rate on $\mathcal N$ changes when the learner is restricted to number-conserving protocols.
\item[(ii)] Let $0<\eps\le1/3600$, $m\ge1$ and $k\ge1$. Every $k$-copy protocol that learns $\mathcal N$ to trace distance $\eps$ with probability at least $2/3$ uses $N>m^3/(8{,}553{,}600\,k\,\eps^2)$ copies, and every protocol whatsoever uses $N>m^2/(8{,}553{,}600\,\eps^2)$ copies. The hard family is $\{\rho_\theta:\theta\in\mathfrak u,\ \|\theta\|_2\le120\sqrt3\,\eps\}$, which lies in $\mathcal N\cap\Gc$ for every $c\le1-\tanh(120\sqrt3\,\eps)$.
\item[(iii)] Hence the collective rate on $\mathcal N$ is $\Theta(m^2/\eps^2)$ (upper bound \cite[Thm.~4.2]{CFGLMWW26}, lower bound by (ii)), and for fixed $c$ and $\eps\le\min\{1/3600,(1-c)/(120\sqrt3)\}$ the single-copy rate on $\mathcal N\cap\Gc$ is $\Theta(m^3/\eps^2)$ (upper bound \cite[Thm.~7.6]{Rub26}, lower bound by (ii)), in both cases whether or not the protocol is required to be number-conserving.
\end{enumerate}
\end{proposition}

\begin{proof}
(i) Consider one round of the protocol, acting on $t$ copies (all $N$ of them for a collective protocol) and on ancillas in a $\theta$-independent state $\eta$, with effects $E_x$. Let $P_t=P^{\otimes t}$ be the total parity of the copies and put $E'_x=\tfrac12\big(E_x+(P_t\otimes I)E_x(P_t\otimes I)\big)$. The $E'_x$ are positive, sum to $I$, and commute with $P_t\otimes I$ because $P_t^2=I$; and for $\rho\in\mathcal S$,
\[
\Tr\big[(\rho^{\otimes t}\otimes\eta)(P_t\otimes I)E_x(P_t\otimes I)\big]=\Tr\big[\big((P_t\rho^{\otimes t}P_t)\otimes\eta\big)E_x\big]=\Tr\big[(\rho^{\otimes t}\otimes\eta)E_x\big],
\]
so the outcome law of the round is unchanged. The protocol's later choices depend only on classical outcomes, so the replacement can be made in every round, and the output has the same law. For particle number let $\hat N_t=\sum_{c=1}^t\hat N^{(c)}$ with spectral projections $\Pi_n$, $0\le n\le mt$, and put $E'_x=\sum_n(\Pi_n\otimes I)E_x(\Pi_n\otimes I)$; the $E'_x$ are positive, sum to $I$, commute with $\hat N_t\otimes I$, and give the same outcome law because $\rho^{\otimes t}$ commutes with every $\Pi_n$. (If the ancillas are themselves fermionic modes prepared in an even, or number-conserving, state, the same identities hold with the total parity, or number, of copies and ancillas together.) A lower bound for unrestricted learners is a lower bound for restricted ones, and every protocol attaining an upper bound can be replaced as above, so the rates are unchanged.

(ii) Let $Q\in\R^{p\times m^2}$ have the orthonormal basis of Lemma~\ref{lem:numberconserving} as columns and parametrise $\mathcal N$ by $\phi\in\R^{m^2}$, $\theta=Q\phi$, so that $\|\theta\|_2=\|\phi\|_2$. By the chain rule the Fisher matrix of any outcome law with respect to $\phi$ is $Q^TI(\theta)Q$, whose trace is $\tr(I(\theta)QQ^T)\le\tr I(\theta)$ because $I(\theta)\succeq0$ and $QQ^T$ is an orthogonal projection. Hence every budget used in the proofs of Theorems~\ref{thm:kcopy-mixed} and~\ref{thm:collective} holds for the $\phi$-family on $\|\phi\|_2\le r_0$: $2mt^2$ for a block of $t$ copies and $2mkn$ for a $k$-copy protocol with $n$ copies (Lemma~\ref{lem:blockfisher}), and $np$ for a collective protocol (Lemma~\ref{lem:qfi}). Set $r=120\sqrt3\,\eps\le r_0$ and place the prior of \cite[Lemma~6.1]{Rub26} on the cube $[-a,a]^{m^2}$ in $\phi$ with $a=r/m$; its Fisher-information trace is $\pi^2m^4/r^2$, and every point of the cube has $\|\theta\|_2\le r$, so the metric comparison \cite[Prop.~5.4]{Rub26} applies on it, and the amplification lemma \cite[Lemma~6.3]{Rub26} applies with the projection taken onto this cube (its proof uses only that the cube is closed, bounded and contained in the ball on which the comparison holds), giving an estimator $\hat\phi$ from $n=99N$ copies with risk at most $0.08\,r^2$, exactly as in the proof of Theorem~\ref{thm:kcopy-mixed}. The van Trees inequality \cite[Lemma~6.2]{Rub26} in dimension $m^2$ gives, for the $k$-copy protocol, $\E\|\hat\phi-\phi\|_2^2\ge m^4/(2mkn+\pi^2m^4/r^2)\ge r^2/(1+\pi^2)>0.09\,r^2$ whenever $n\le m^3/(2kr^2)$. Hence $99N>m^3/(2kr^2)=m^3/(86{,}400\,k\,\eps^2)$, that is $N>m^3/(8{,}553{,}600\,k\,\eps^2)$. For a collective protocol the same inequality with the budget $np$ gives risk larger than $0.09\,r^2$ whenever $n\le m^4/(pr^2)$, so $99N>m^4/(pr^2)\ge m^2/(2r^2)=m^2/(86{,}400\,\eps^2)$ using $p\le2m^2$, that is $N>m^2/(8{,}553{,}600\,\eps^2)$. Every state in play is $\rho_\theta$ with $\theta\in\mathfrak u$ and $\|\theta\|_2\le r$; its normal-form values are at most $\tanh r$, which gives the last statement.

(iii) The upper bounds hold on every subclass of the classes on which they are proved, and by (i) they are attained by number-conserving measurements on $\mathcal N$.
\end{proof}

\begin{remark}
The tomographic analogue of the question raised in \cite{MSR26} for dynamics therefore has a complete answer on the number-conserving Gaussian states: number superselection costs nothing, and entangled measurements remain necessary for the optimal rate. (The dynamical question has an obstruction with no tomographic counterpart: by \cite[Rem.~5.2]{MSR26}, some parity-preserving but number-nonconserving access is necessary for full identification of the generator.) The hardness of Theorems~\ref{thm:kcopy-mixed} and~\ref{thm:collective} lives entirely inside the number-conserving slice, which contains the finite-temperature states of every free-fermion Hamiltonian that conserves particle number. On states that do not conserve particle number the question changes character, since a number-conserving measurement sees only the pinched state $\sum_n\Pi_n\rho\Pi_n$, which need not be Gaussian, and it is not pursued here.
\end{remark}

\section{Open questions}\label{sec:open}

\begin{question}[$k$-copy upper bounds]\label{q:upper}
Theorems~\ref{cor:kcopy-rate} and~\ref{thm:cold-general} determine the $k$-copy rate on the cold number-conserving classes and on the cold bosonic class for every $k$. Is $m^3/(\eps^2\min\{\sqrt k,m\})$ attained by $k$-copy protocols on all mixed fermionic Gaussian states with covariance a fixed distance from the pure boundary, where the lower bound is only the $1/k$ of Theorem~\ref{thm:kcopy-mixed}, or on the fermionic Gaussian states near the vacuum with pairing terms, where there is no heterodyne stage to remove them? And what is the $k$-copy rate at intermediate temperatures, $1/m\ll\nu\ll1$, where the single-copy rate is $m^2/(\nu\eps^2)$ \cite{CMMP26} and nothing beyond the trivial $\Omega(m^2/\eps^2)$ and $O(m^2/(\nu\eps^2))$ is known for $k>1$? The Bayesian method of \cite[Sec.~5]{CMMP26} is single-copy, and the embedding does not reach warm states: the symmetriser accepts $t$ copies of a qudit near $I/m$ with probability $\binom{m+t-1}{t}m^{-t}$, between $1/t!$ and $e^{t^2/(2m)}/t!$, so a state with $\Theta(m)$ photons cannot be sampled from qudit copies at polynomial cost, which is the difficulty \cite{CMMP26} describe for the random purification channel.
\end{question}

\begin{question}[the $k$-copy rate at the maximally mixed state]
Theorems~\ref{thm:exact} and~\ref{thm:asymptotic} show that the block Fisher budget of Lemma~\ref{lem:blockfisher} cannot be improved by more than a constant factor for any $t\le m/32$, so the $1/k$ of Theorem~\ref{thm:kcopy-mixed} is the best that the uniform trace-budget route can give there. Is it the truth? That is, is there a $k$-copy protocol learning mixed fermionic Gaussian states near $I/2^m$ with $O(m^3/(k\eps^2))$ copies for $k\le m$? The covariant coherent-state measurement of Theorem~\ref{thm:exact} has the right Fisher information; what is missing is an estimator that attains it, and an extension of Theorems~\ref{thm:exact} and~\ref{thm:asymptotic} from $\theta=0$ to the ball $\|\theta\|_2\le r_0$. A positive answer would make the maximally mixed corner strictly easier for few-copy measurements than the vacuum, where the rate is $m^3/\sqrt k$.
\end{question}

\begin{question}[the fermionic crossover in $c$]
On $\Gc$ the single-copy rate is $\Theta(m^3/\eps^2)$ for fixed $c$, with the upper bound $O(m^3/(c\eps^2))$ \cite[Thm.~7.6]{Rub26}. Theorems~\ref{thm:vacuum} and~\ref{cor:kcopy-rate} show that on the number-conserving class with bounded fugacity trace, which contains a hard local family with covariance gap of order $1/m$, the single-copy rate is still $\Theta(m^3/\eps^2)$, so a loss of $1/c$ is not necessary on that subclass; on the full class $\Gc$ with $c$ decreasing in $m$ the dependence on $c$ is open. Is the rate $\tilde\Theta(m^3/\eps^2)$ uniformly in $c$ (Conjecture~10.1 of \cite{Rub26}), or is there an intermediate regime? The Bayesian anti-concentration method of \cite[Sec.~5]{CMMP26}, which needs no embedding, is the natural tool for a $c$-dependent lower bound.
\end{question}

\begin{question}[testing]
Proposition~\ref{prop:transfer} moves block-model lower bounds. The single-copy lower bounds for mixedness testing and state certification of unstructured states \cite{BCL20,CHLL22} are stated for single-copy measurements, which the sampler does not preserve, since one Gaussian copy becomes a random block of qudit copies. A weighted block-model version of those bounds would give, through Proposition~\ref{prop:transfer}, single-copy lower bounds for testing whether a near-vacuum fermionic or bosonic Gaussian state equals the isotropic thermal state $\rho^{\rm F}_{I/m}$; Question~10.6 of \cite{Rub26} asks for such bounds.
\end{question}

\section*{Acknowledgements}
I thank Senrui Chen, Antonio Anna Mele, Francesco Anna Mele and John Preskill for correspondence about the relation between \cite{CMMP26} and \cite{Rub26}.

\section*{Statement on the use of artificial-intelligence tools}
The results in this paper were obtained by the author. LLM systems were used to check work and help with calculations and to prepare the manuscript text, as well as search the literature and check references. No language model is an author, and the author takes full responsibility for every statement in this paper.

\end{document}